\documentclass[twocolumn]{revtex4}
\usepackage[dvipdfmx]{graphicx}
\usepackage{amsmath,amsbsy,amssymb}
\usepackage{bm}
\usepackage{mathrsfs}
\usepackage{ulem}
\usepackage{physics}
\usepackage{textgreek}
\usepackage{mathrsfs}
\usepackage{multirow}
\usepackage{slashed}

\usepackage{color}

\newcommand{\diff}{\mathrm{d}}

\newcommand{\imu}{\mathrm{i}}
\newcommand{\epn}{\mathrm{e}}

\newcommand{\ua}{\uparrow}
\newcommand{\da}{\downarrow}
\newcommand{\dg}{\dagger}
\newcommand{\la}{\langle}
\newcommand{\ra}{\rangle}
\newcommand{\al}{\alpha}
\newcommand{\sg}{\sigma}
\newcommand{\gm}{\gamma}
\newcommand{\ep}{\varepsilon}

\newcommand{\dvec}[1]{\hspace{-1mm}\stackrel{\leftrightarrow}{#1}\hspace{-1mm}}

\newcommand{\ldvec}[1]{\hspace{-1mm}\stackrel{\longleftrightarrow}{#1}\hspace{-1mm}}

\begin{document}

\title{
Dirac bilinears
in condensed matter physics:
\\
Relativistic correction
for observables
and conjugate electromagnetic fields
}

\author{
Shintaro Hoshino$^1$, Tatsuya Miki$^1$, Michi-To Suzuki$^{2, 4}$, and Hiroaki Ikeda$^3$
}

\affiliation{
$^1$Department of Physics, Saitama University, Sakura, Saitama 338-8570, Japan 
\\
$^2$Department of Materials Science, Graduate School of Engineering, Osaka Metropolitan University, Sakai, Osaka 599-8531, Japan
\\
$^3$Department of Physics, Ritsumeikan University, Kusatsu, Shiga 525-8577, Japan
\\
$^4$Center for Spintronics Research Network, Graduate School of Engineering Science, Osaka University, Toyonaka, Osaka 560-8531, Japan
}

\date{\today}

\begin{abstract}

Inspired by recent developments in electron chirality, we reconsider some microscopic physical quantities that have been overlooked or have received little attention in condensed matter physics, based on the non-relativistic limit of the Dirac bilinears in relativistic quantum theory. 
We identify the expression 
of physical quantities defined by the four-component Dirac field
in terms of
the two-component Schr\"odinger field, which is usually used in condensed matter physics, and clarify its conjugate electromagnetic field. 
This consideration bridges the fields of condensed matter physics, quantum chemistry, and particle physics, and paves the way to electromagnetic control of matter. 
Our findings provide a means of {\it ab initio} quantification of material 
characters such as chirality and axiality that are unique to low-symmetry materials, and 
stimulate 
the systematic search for useful, new functionalities.

\end{abstract}

\maketitle

\section{Introduction}

The diversity of quantum phenomena in solids is caused by the low symmetry of the electronic states, which is a characteristic feature for condensed matter physics.
Specifically, the crystals with 
polar, chiral, and (ferro-)axial 
properties have recently attracted considerable attention due to their intriguing electronic phenomena and functionalities \cite{Tokura18,Tokura21,Spaldin08,Resta94,Vanderbilt_book,Hayashida20,Zhou20,Inui20,Gohler11,Nakajima15}.
Theoretically, these behaviors should be characterized quantitatively by the spatiotemporal distributions of fundamental microscopic physical quantities such as charge, spin, and current densities.

Recently, systematic research for condensed matter using the concept of multipole moments has been actively pursued, and
especially, electric toroidal multipoles, which have not been widely recognized, are attracting attention as a novel non-magnetic degree of freedom \cite{Dubovik86,Hayami18,Hayami19,Hirose22,Oiwa22,Hayami22,Kishine22}.
In our previous study \cite{Hoshino23}, from the specific physical description of the multipole moments of localized atomic orbitals, it becomes apparent that the electric toroidal multipole moments cannot be fully captured by fundamental physical quantities such as charge, spin, and current density, necessitating a reconsideration of microscopic quantities.
With the knowledge of relativistic quantum mechanics, the basic physical quantities are elucidated, and associated with electric polarization derived from spin degrees of freedom ($\bm P_S$) and electron chirality ($\gm^5$).
The spatial distribution of these quantities have been discussed also in the context of spintronics \cite{Wang06,Rodina08} and quantum chemistry \cite{Bast11,Tachibana12,Hara12,Fukuda16, Senami18,Senami19, Senami20,Kuroda22,Kuroda23}.

\begin{figure*}[t]
\begin{center}
\includegraphics[width=130mm]{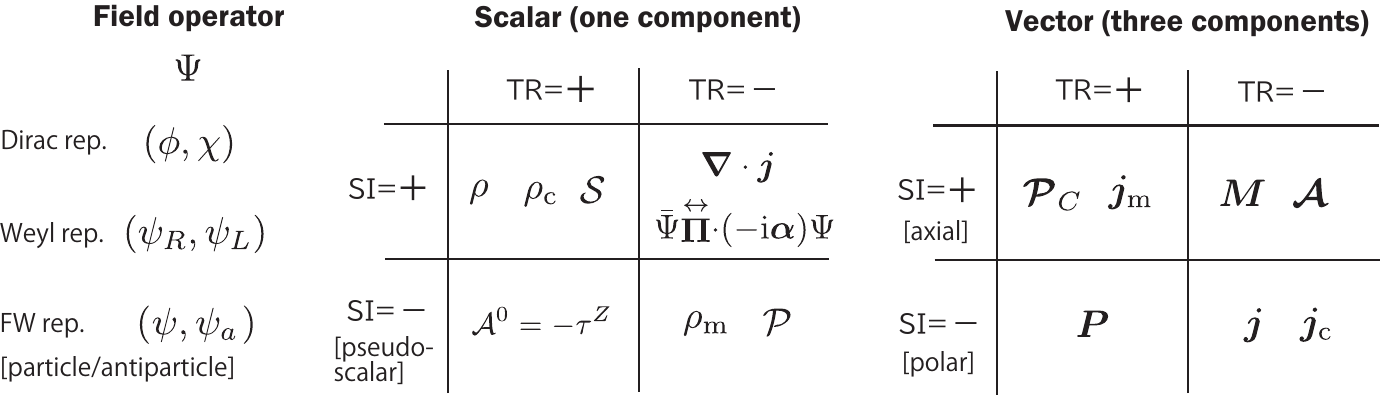}
\caption{
Microscopic physical quantities considered in this paper.
They are classified in terms of scalar (one component) or vector (three components) as a transformation property in the three-dimensional space.
The signs from spatial inversion $\mathscr P$ (SI) and time reversal $\mathscr T$ (TR) for a quantity $O(\bm r)$ are defined by 
$\mathscr P O(\bm r) \mathscr P^{-1} = \pm O(-\bm r)$
and 
$\mathscr T O(\bm r) \mathscr T^{-1} = \pm O(\bm r)$
, respectively, where
$\mathscr T$ is an antiunitary operator.
The pseudoscalar and polar/axial vectors in this table are defined based on spatial inversion.
Note that the terminology in the figure is different from relativistic quantum mechanics based on Lorentz transformation.
}
\label{fig:summary}
\end{center}
\end{figure*}

In this paper, we provide a 
full
set of the microscopic physical quantities (Dirac bilinears) in condensed matter physics based on relativistic quantum mechanics, which is useful to quantify the low symmetry of materials.
Here ``microscopic'' means that the physical quantity is defined in terms of four-component Dirac field at every spatiotemperal point together with the gauge invariance and Hermiticity.
Everything is contained within four-component Dirac field description, making it difficult to understand, however.
For the physical interpretations of the results, it is suitable to take the non-relativistic limit (NRL) where the antiparticles are absent.
Relativistic correction for physical quantities are not typically considered so far, but in the field of condensed matter, there are various overlooked physical quantities with 
interesting effects and significance. 
The examples include electron chirality $\gm^5$ and electric polarization $\bm P$ \cite{Hoshino23}.
Here, we derive many other quantities (see Fig.~\ref{fig:summary}) and their relativistic corrections, which have a potential to characterize the property of materials
such as 
polarity, chirality and axiality, leading to a systematic search for the useful functionalities of condensed matter \cite{Miki24}.

This research lies at the interdisciplinary frontier of solid state physics, quantum chemistry, and particle physics. 
The knowledge should be shared among all of these fields.
The essence of this paper is all included in the beautifully symmetric Dirac equation $(\imu \slashed{D}-m)\Psi = 0$, from which complicated Hamiltonian and physical quantities in solids emerge by removing the antiparticle sector.
The correspondence between four-component Dirac-field representation and two-component Schr\"odinger field representation, which is discussed in this paper in detail, serves as a dictionary of elementary particle physics and condensed matter physics.

Previously, we do not consider the external field other than the crystalline electric field \cite{Hoshino23}.
Here we allow the time-dependence under the arbitrary electromagnetic (EM) fields, and clarifies the coupling between electron and external fields.

This paper is organized as follows.
In the next section, the microscopic physical quantities in terms of the Dirac field $\Psi$ are summarized based on relativistic quantum mechanics.
The relativistic corrections for physical quantities are given in terms of the Schr\"odinger field $\psi$ in Sec. III.
We reconsider the Bloch-Bohm theorem for two kinds of electronic currents in Sec. IV.
A summary of this paper is provided in Sec. V.
Appendix A provides a derivation of quantum anomaly in statistical mechanics for reference.
A simple electron gas model for chiral and polar metals is discussed in Appendix B.
Appendix C is devoted to the high-frequency expansion of effective Hamiltonian under the fast oscillating external fields without taking NRL.

\section{Physical quantities in terms of Dirac field}

Here, we re-examine physical quantities using the Dirac equation. 
Through this analysis, we identify several useful relationships, which 
suggest previously overlooked physical quantities in condensed matter physics.
These quantities should be observable with current advanced technology. 
The following sections will systematically organize these findings.

\subsection{Hamiltonian}
We begin with the Hamiltonian in relativistic quantum mechanics for electrons.
The Dirac equation is given by
\begin{align}
    &\imu \hbar \partial_t \Psi = H \Psi
    , \label{eq:Dirac_Equation}
    \\
    &H = c \bm \al \cdot \bm \Pi + \beta mc^2 
    + e\Phi
    ,
    \\
    &\bm \al = \begin{pmatrix}
        0 & \bm \sg \\
        \bm \sg & 0
    \end{pmatrix}
    ,\ \ 
    \beta = \begin{pmatrix}
        1 & 0 \\
        0 & -1
    \end{pmatrix}
    ,
\end{align}
where 
$\Psi$ is the four component Dirac spinor field 
in Dirac (or standard \cite{Berestetskii_book}) representation, and $\beta,\bm \al$ are the $4\times 4$ matrices.
The $2\times 2$ Pauli matrices are written as $\bm \sg$.
The canonical momentum is given by $\bm \Pi = \bm p -\frac e c \bm A$
with $\bm p= -\imu\hbar \bm \nabla$.

The expectation value of the electronic Hamiltonian, i.e. energy, is given by
\begin{align}
    \mathscr H&= \int \diff \bm r \Psi^\dg H \Psi
    .
\end{align}
In the second quantization formalism, the field $\Psi$ is promoted to an operator to describe many-particle systems (quantum field theory).
In the following, we basically treat $\Psi$ as a complex 
field unless stated otherwise.

We also introduce the action $\mathscr S = \int \diff t \mathscr L$
where the Lagrangian is given by
\begin{align}
    \mathscr L &= \int \diff \bm r \Psi^\dg \qty(\imu \hbar \partial_t - H )\Psi
    + \frac{1}{8\pi}  \int \diff \bm r (\bm E^2 - \bm B^2)
    .\label{eq:gen_Lagrangian}
\end{align}
The EM fields are given $\bm E= -\bm \nabla \Phi - \frac 1 c \frac{\partial \bm A}{\partial t}$ and $\bm B = \bm\nabla \times \bm A$.
The equation of motion is derived from the least principle of the action, to result in the Dirac equation and Maxwell equation.
We identify the electric charge density $\rho$ and current density $\bm j$ as the source term in the Maxwell equation \cite{Wang06}.
In the next subsection, we list a number of microscopic physical quantities in addition to $\rho$ and $\bm j$, which are defined in terms of Dirac field at every spatiotemporal point.
The EM fields are treated as classical throughout the paper, and are regarded as external fields.

In the following, both the $\gamma^\mu$-representation and $(\beta,\bm \al)$-representations are used.
While the former is suitable for recognizing the symmetry property and the structure of the theory, the latter is convenient for the actual calculations in NRL and for relativistic corrections.

\subsection{Dirac bilinears}

Here, we summarize the Dirac bilinears and their relations.
First of all, we introduce the gamma matrices by $\gm^0 = \beta$ and $\gm^{1,2,3} = \beta\al^{x,y,z}$.
We also define the matrices by \cite{Berestetskii_book}
\begin{align}
    \gm^5 &= -\imu \gm^0 \gm^1 \gm^2 \gm^3
    = \begin{pmatrix}
        0 & -1 \\
        -1 & 0
    \end{pmatrix}
    , \\
    \bm \Sigma &= -\bm \al \gm^5
    = \begin{pmatrix}
        \bm \sg & 0 \\
        0 & \bm \sg
    \end{pmatrix}
    ,
\end{align}
and antisymmetric tensor by $\sg^{\mu\nu} = \frac 1 2 [\gm^\mu, \gm^\nu]$ ($\mu,\nu = 0,1,2,3$).
Since the Dirac spinor $\Psi$ is given by four-component vector, 
there are in total $4\times 4 = 16$ (=1+1+4+4+6) components of the Dirac bilinears.
They are classified based on Lorentz transformation as \cite{Berestetskii_book}
\begin{align}
    \mathcal S &= \bar \Psi \Psi
    , \label{eq:DB_S}
    \\
    \mathcal P &= \imu \bar \Psi \gm^5 \Psi
    , \label{eq:DB_P}
    \\
    \mathcal V^\mu &= \bar \Psi \gm^\mu \Psi
    = (\mathcal V^0 , \bm {\mathcal V})
    , \label{eq:DB_V}
    \\
    \mathcal A^\mu &= \bar \Psi \gm ^\mu \gm^5 \Psi
    = (\mathcal A^0 , \bm {\mathcal A})
    , \label{eq:DB_A}
    \\
    \mathcal T^{\mu\nu} &= \imu \bar \Psi \sg^{\mu\nu} \Psi
    = \imu \bar\Psi \big( \bm \al, \imu \bm \Sigma \big) \Psi
    ,\label{eq:DB_T}
\end{align}
each of which corresponds to the Lorentz scalar, pseudoscalar, vector (or four-component current), pseudovector (or four-component axial current), and tensor.
We have used the same notation as Ref.~\cite{Berestetskii_book} for Dirac bilinears.
$\bar \Psi = \Psi^\dg \gm^0$ is the Dirac conjugate.
All of these quantities are real-valued (or Hermitian in second-quantized form).
The antisymmetric tensor $\mathcal T$ is represented by the pair of two three-dimensional vectors $ \bm \al$ and $\bm \Sigma$, whose explicit representation is given by
\begin{align}
\big( \bm \al, \imu \bm \Sigma \big)
&= 
\begin{pmatrix}
     0 & \al^x & \al^y & \al^z \\
     -\al^x & 0 & -\imu \Sigma^z & \imu \Sigma^y \\
     -\al^y & \imu \Sigma^z & 0 & -\imu \Sigma^x \\
     -\al^z & -\imu \Sigma^y & \imu \Sigma^x & 0 \\
 \end{pmatrix} 
 .
\end{align}
Similarly, using this notation, 
the antisymmetric EM tensor is written in terms of the electric and magnetic fields as $F_{\mu\nu} = \partial_\mu A_\nu - \partial_\nu A_\mu = (\bm E,\bm B)$ and $F^{\mu\nu} = (-\bm E,\bm B)$, where $A^\mu =(\Phi,\bm A)$ and $A_\mu = (\Phi,-\bm A)$ \cite{Berestetskii_book}.
The above quantities are obviously U(1) gauge-invariant, i.e., invariant under the phase transformation $\Psi \to \epn^{\imu \theta} \Psi$.

The sixteen 4$\times$4 matrices satisfy the orthonormality condition and are complete \cite{Berestetskii_book}.
Namely, when we write the Dirac bilinears as $\mathcal O^\xi = \Psi^\dg O^\xi \Psi$, the $4\times 4$ matrix $O^\xi$ ($\xi=1,\cdots 16$) satisfies
\begin{align}
&O^{\xi\dg} = O^\xi = (O^\xi)^{-1}
, \\
&{\rm Tr\,} O^\xi O^{\xi'} = 4\delta_{\xi\xi'}
, \\
&\sum_{\xi=1}^{16} O^{\xi*}_{ij} O^{\xi}_{kl} = 4\delta_{ik} \delta_{jl}
, \\
& \det O^\xi = 1
\end{align}
We also define $A = \beta O$, where the matrix $O$ is chosen as one of $\{O^{\xi}\}$.
The relation $A^\dg = \gm^0 A \gm^0$ is satisfied due to $O=O^\dg$.
Note that $A$ is not necessarily Hermitian matrix, but the quantity $\bar \Psi A \Psi$ is real-valued.
Using the Dirac equation, the quantities characterized by a matrix $A$ can be rewritten as 
\begin{align}
    \bar \Psi A \Psi &= \frac{1}{4mc^2} \Big[
    \hbar \bar \Psi (\ldvec{\ \imu D_t\ }) \{ A,\gm^0\} \Psi 
    + \imu \hbar \partial_t \qty( \bar \Psi [A,\gm^0] \Psi)
\nonumber \\
&\hspace{5mm}
- c\bar \Psi \dvec{\bm \Pi } \cdot \{ A,\bm \gm \} \Psi 
    + \imu \hbar c \bm \nabla \cdot  \qty( \bar \Psi [A,\bm \gm] \Psi)
\Big]
. \label{eq:gordon_decomp_gen}
\end{align}
We have defined the symbol $A \dvec{O} B = A O B + (O^*A) B$: for example, $\Psi^\dg \dvec{\bm p} \Psi = \Psi^\dg \bm p \Psi - (\bm p \Psi^\dg)\Psi$. 
The canonical time-derivative is given by $D_t = \partial_t + \imu \frac{e}{\hbar} \Phi$.
The application of the relation \eqref{eq:gordon_decomp_gen} to charge and current
is known as the Gordon decomposition \cite{Gordon28, Baym_book}.
It is also convenient to write down the equation of motion for Dirac bilinears explicitly as follows.
\begin{align}
    &\hbar \partial_t (\Psi^\dg O\Psi)  + \hbar c \bm \nabla \cdot \qty(\Psi^\dg\frac 1 2 \{O,\bm \al \}  \Psi)
    \nonumber \\
    &\hspace{5mm} =
    c \Psi^\dg \frac 1 {2\imu} [ O,\bm \al ]  \cdot \dvec{\bm \Pi}\Psi 
    + mc^2 \Psi^\dg\frac 1 \imu [ O,\beta ] \Psi
    . \label{eq:time_derivative_formula}
\end{align}
The right-hand side of Eq.~\eqref{eq:time_derivative_formula} is interpreted as a source or sink of the physical quantity $\Psi^\dg O\Psi$.
We note that the above two equations are obtained by using the Dirac equation (so-called ``on-shell'' condition), and are useful to clarify the relations among physical quantities.

\subsection{Relations among Dirac bilinears}

In this subsection, we take a closer look at each Dirac bilinear and their relations.
Since we are interested in expressions in NRL, which will be discussed in Sec. III, we use a representation suitable for taking the limit $c\to \infty$ or $m\to \infty$.
Namely, we write the Dirac field as
\begin{align}
\Psi =  \begin{pmatrix}
    \phi \\ \chi
\end{pmatrix}
\epn^{-\imu mc^2 t/\hbar}
, \label{eq:def_of_phi_chi}
\end{align}
which is composed of two-component spinors $\phi$ and $\chi$ (Dirac representation).
We have shifted the energy by $mc^2$ for convenience when taking NRL later.

\subsubsection{Lorentz vector (\,$\mathcal V^\mu$) and tensor  (\,$\mathcal T^{\mu\nu}$)}

Let us consider the four-component vector $\mathcal V^\mu = (\mathcal V^0,\bm {\mathcal V})$; the time-component is the charge density and the spatial component corresponds to the current density.
We define the quantities
\begin{align}
    \rho &=e\mathcal V^0 = e\bar \Psi \gm^0 \Psi = e\Psi^\dg \Psi = e(\phi^\dg \phi + \chi^\dg \chi)
    , \\
    \bm j &=ec\bm {\mathcal V} = ec \bar \Psi \bm \gm \Psi = ec\Psi^\dg \bm \al\Psi = ec (\phi^\dg \bm \sg \chi + {\rm conj.})
    ,
\end{align}
which appear as a source term in the Maxwell equation.
The three component vector $c\bm \al$ is regarded as a velocity, which is helpful to understand the physical meaning of each matrix.

To gain more insight, we further rewrite them using the Gordon decomposition \cite{Gordon28, Baym_book}:
\begin{align}
\rho &= \rho_{\rm c} - \bm \nabla \cdot \bm P
, \label{eq:gordon_charge}
\\
    \bm j &= \bm j_{\rm c} + c\bm \nabla \times \bm M + \frac{\partial \bm P}{\partial t}
    , \label{eq:gordon_current}
\end{align}
where we have introduced the `convective' charge and current as \cite{Baym_book}
\begin{align}
    \rho_{\rm c} &= \frac{\hbar e}{2mc^2} \bar \Psi \ldvec{(\imu D_t)} \Psi 
    , \label{eq:convective_charge}
    \\
    \bm j_{\rm c} &= \frac{e}{2m} \bar \Psi \dvec{\bm \Pi} \Psi 
    . \label{eq:convective_current}
\end{align}
We have also defined the {\it microscopic} magnetization and electric polarization by
\begin{align}
    \bm M &= \frac{\hbar e}{2mc} \bar \Psi \bm \Sigma \Psi
    = \frac{\hbar e}{2mc} (\phi^\dg \bm \sg \phi - \chi^\dg \bm \sg\chi)
    , \label{eq:def_of_M}
    \\
    \bm P &= \frac{\hbar e}{2mc} \bar \Psi (-\imu \bm \al) \Psi
    = -\frac{\imu \hbar e}{2mc} (\phi^\dg \bm \sg \chi - {\rm conj.})
    , \label{eq:def_of_P}
\end{align}
which are clearly related to the tensor components as $\mathcal T^{\mu\nu} = - \frac{2mc }{\hbar e} (\bm P,\bm M)$ $\equiv (\bm {\mathcal P}, \bm {\mathcal M})$ given in Eq.~\eqref{eq:DB_T}.
Whereas usually $\sg^{\mu\nu}$ is recognized as a generator of Lorentz transformation, which defines the Lorentz scalar, we can also interpret the tensor as physical quantities of electric polarization and magnetization.
The equation of motion for charge density results in \cite{Baym_book}:
\begin{align}
    \frac{\partial \rho}{\partial t} + \bm \nabla \cdot \bm j 
    = \frac{\partial \rho_{\rm c}}{\partial t} + \bm \nabla \cdot \bm j_{\rm c}
    = 0
    . \label{eq:eoc_charge_Dirac}
\end{align}
This is a conservation law of charges.

Using the equation of motion for the electric polarization [i.e., Eq.~\eqref{eq:gordon_current}], we obtain the orbital angular momentum:
\begin{align}
    &mc \int \diff \bm r  \qty(\bm r\times \Psi^\dg \bm \al \Psi)
    \nonumber \\
    &= \int \diff \bm r 
    \bar \Psi \qty( \bm L + \hbar \bm \Sigma ) \Psi
    + \frac{m}{e}\int \diff \bm r \qty( \bm r\times \frac{\partial \bm P}{\partial t} )
    ,
\end{align}
where we have used the integration by parts and dropped the surface term.
$\bm L = \bm r\times \bm \Pi $ is the angular momentum operator.
In the stationary case, the time-derivative of electric polarization vanishes.
Then, noting that the velocity is given by $\bm v (\sim \dot {\bm r}) =c\bm \al$, the above equation may be symbolically written as $\bm r\times m \bm v = \bm L + 2\bm S$ with the spin angular momentum $\bm S = \frac{\hbar }{2} \bm \Sigma$, which is related to the total 
magnetic moment 
\cite{Dirac_book}.

We also consider the equation of motion for the electric current density:
\begin{align}
    \frac{\partial \bm j}{\partial t}
    &= \frac{ec^2}{\hbar} \Psi^\dg \dvec{\bm \Pi}\times \bm\Sigma \Psi
    - c^2 \bm \nabla \rho - \frac{4m^2c^4}{\hbar^2} \bm P
    .\label{eq:cur_t_deriv}
\end{align}
Notably, as shown in the right-hand side of this equation, the time derivative of current is related to the electric polarizations, where $\bm P$ defined in Eq.~\eqref{eq:def_of_P} is one of such contributions.
We will utilize this technique for defining the ``chirality polarization'' by considering the time-derivative of axial current, as shown in the next subsection [see Eq.~\eqref{eq:def_chiral_polarization}].

\subsubsection{Lorentz pseudovector (\,$\mathcal A^\mu$) and pseudoscalar (\,$\mathcal P$)}

Now we consider the four-component pseudovector $\mathcal A^\mu = (\mathcal A^0, \bm {\mathcal A})$.
The explicit forms are
\begin{align}
    \mathcal A^0 &= \Psi^\dg \gm^5 \Psi = - (\phi^\dg \chi + {\rm conj.})
    , \label{eq:Dirac_chirality}
    \\
    \bm {\mathcal A} &= - \Psi^\dg \bm \Sigma \Psi
    =  - (\phi^\dg \bm \sg \phi + \chi^\dg \bm \sg\chi)
    . \label{eq:Dirac_chirality_current}
\end{align}
The time-component is known as the chirality density.
The spatial part is regarded as the (three-component) axial current, which corresponds to a current of the chirality density.
The Gordon decomposition of these quantities provides a relation to the other Dirac bilinears as
\begin{align}
    2mc^2 \mathcal A^0 &= - c\bar \Psi \dvec{\bm \Pi} \cdot \bm \Sigma \Psi 
    - \hbar\frac{\partial \mathcal P}{\partial t}
    , \label{eq:A0_Gordon}
    \\
    2mc^2 \bm {\mathcal A} &= - \hbar \bar \Psi \ldvec{(\imu D_t)} \bm \Sigma \Psi - c \bar \Psi \dvec{\bm \Pi} \times (-\imu \bm \al) \Psi 
    + \hbar c \bm \nabla \mathcal P
    . \label{eq:Avec_Gordon}
\end{align}
The right-hand sides of Eqs.~\eqref{eq:A0_Gordon} and \eqref{eq:Avec_Gordon}
give another expression of electron chirality and axial current, respectively.
For example, in Eq.~\eqref{eq:A0_Gordon}, the time-derivative of pseudoscalar gives a contribution to the chirality.
Another contribution to chirality, $\bm \Sigma \cdot \bm \Pi$, is an inner product between 
magnetization and momentum, which can be regarded as helicity density.
In the stationary state, the time derivative vanishes ($\partial \mathcal P / \partial t = 0$), and hence
$\mathcal A^0$ is identical to $-\bar \Psi \dvec{\bm \Pi} \cdot \bm \Sigma \Psi / 2mc$.
It is also interesting to mention another form of the chirality density:
\begin{align}
    \mathcal A^0 = - \frac 1 {3c}  \Psi^\dg \bm \Sigma \cdot \bm  v \Psi
    . \label{eq:Another_rep_of_A0}
\end{align}
Namely, the chirality is given by the inner product of the spin angular momentum and velocity
\cite{Kuroda24}, which is again interpreted as a kind of helicity.
In this way, the rewriting of the abstract quantity $\gm^5$ helps us to understand its physical meaning.

The pseudoscalar $\mathcal P$ acts as a potential for the chirality,
since $\partial_t \mathcal P$ and $\bm\nabla \mathcal P$ respectively contribute to electron chirality and axial current as shown in Eqs.~\eqref{eq:A0_Gordon} and \eqref{eq:Avec_Gordon}.
Let us consider this quantity in more detail:
\begin{align}
    \mathcal P &= \Psi^\dg \imu \beta \gm^5 \Psi = \Psi^\dg \gm^1\gm^2\gm^3 \Psi = -\imu (\phi^\dg \chi - {\rm conj.})
    .
\end{align}
The Gordon decomposition is given by
\begin{align}
    2m c^2 \mathcal P &= \hbar \frac{\partial \mathcal A^0}{\partial t}
    + \hbar c \bm \nabla \cdot \bm {\mathcal A}
    ,
\end{align}
which is identical to the equation of continuity for the pseudovector $\mathcal A^\mu$.
The pseudoscalar $\mathcal P$ is regarded as a source term of the chirality density.
Note that the chiral anomaly term is added to the left-hand side if the quantum field theory is considered \cite{Fujikawa_book, Hoshino23} (see also Sec. II\,F).

With use of Eq.~\eqref{eq:time_derivative_formula}, 
the time evolution of axial current $\bm {\mathcal A}$ is obtained as
\begin{align}
    \frac 1 c \frac{\partial \bm {\mathcal A}}{\partial t}  + \bm \nabla \mathcal A^0 &= - 
    \frac{1}{\hbar} \Psi^\dg \dvec{\bm \Pi}\times \bm \al \Psi
     . \label{eq:Axial_cur_t_deriv}
\end{align}
If we regard $\bm {\mathcal A} = - \Psi^\dg \bm \Sigma \Psi$ as the spin density, $\bm \nabla \mathcal A^0$ is regarded as one of the contributions to the spin torque \cite{Tachibana12,Fukuda16}.
Multiplying the position vector $\bm r$ and integrating this equation over the space, we obtain the total chirality
\begin{align}
    \int \diff \bm r \mathcal A^0 &= \frac 1 {3c} \int \diff \bm r \frac{\partial (\bm r \cdot \bm {\mathcal A}) }{\partial t}  + \frac{2}{3\hbar c}\int \diff \bm r \Psi^\dg \bm L \cdot \bm v \Psi
    .
\end{align}
A similar relation is considered also in Ref.~\cite{Roberts14}.
If we consider the stationary state, the first term on the right-hand side vanishes.
Hence the total chirality is given by the inner product of $\bm L$ and $\bm v$, where spin is not apparently involved in contrast to Eqs.~\eqref{eq:A0_Gordon} and \eqref{eq:Another_rep_of_A0}.

Equation~\eqref{eq:Axial_cur_t_deriv} is compared to the time-derivative of current in Eq.~\eqref{eq:cur_t_deriv}, which defines the electric polarization.
In a similar manner, the right-hand side of Eq.~\eqref{eq:Axial_cur_t_deriv} is regarded as the chirality polarization $\bm {\mathcal P}_C$:
\begin{align}
    \bm {\mathcal P}_C &= \frac{c}{2} \Psi^\dg \dvec{\bm \Pi}\times \bm \al \Psi
    , \label{eq:def_chiral_polarization}
\end{align}
which has a dimension of energy density.
This is interpreted as outer product of momentum and velocity: $\bm \Pi \times \bm v$.
Since the directions of momentum and velocity is same in NRL, the chirality polarization in Eq.~\eqref{eq:def_chiral_polarization} reflects the relativistic effect.
The distribution of the chirality polarization $\bm {\mathcal P}_C$ is expected to characterize a ferroaxial material \cite{Hayashida20}, which is analogous to the quantification of a polar material by the electric polarization $\bm P$.
It is also seen from Eq.~\eqref{eq:Axial_cur_t_deriv} that $\bm \nabla \mathcal A^0$ is regarded as a chirality polarization, and is identical to $-\tfrac{2}{\hbar c}\bm {\mathcal P}_C$ in a stationary state.

To gain more insight into the chirality discussed above, 
it is convenient to introduce the Weyl (or spinor \cite{Berestetskii_book}) basis defined by
\begin{align}
    \psi_R &= (\phi + \chi )/\sqrt 2 ,
    \\
    \psi_L &= (\phi - \chi )/ \sqrt 2 ,
\end{align}
where $R$ and $L$ indicate chirality degrees of freedom of the Dirac spinor.
This representation is suitable for massless particles, where $\psi_{R,L}$ is the eigenfunction of the Dirac Hamiltonian and the helicity operator $h = \bm \sg \cdot \bm p/|\bm p|$ \cite{Gross_book}.
With the Weyl basis ($R,L$), we can define the pseudospin representation in combination with Pauli matrices:
\begin{align}
\tau^0 &= \psi^\dg_R \psi_R + \psi_L^\dg \psi_L = \mathcal V^0
\\
    \tau^Z &= \psi_R^\dg \psi_R - \psi_L^\dg \psi_L = - \mathcal A^0
    , \\
    \tau^X &= \psi_R^\dg \psi_L + \psi_L^\dg \psi_R 
    \ \ = \mathcal S
    , \\
    \tau^Y &= -\imu (\psi_R^\dg \psi_L - \psi_L^\dg \psi_R) = - \mathcal P
    .
\end{align}
The $z$-component $\tau^Z$ in Weyl representation is the chirality density which measures the difference of the number of right- and left-handed electrons.
\\ 
\indent
The chirality density has similarly been considered for EM fields: 
$C = \bm E\cdot (\bm \nabla \times \bm E) + \bm B \cdot (\bm \nabla \times \bm B)$.
This quantity, known as Lipkin's zilch, was initially referred to as ``zilch'' (meaning ``nothing'') due to its unclear physical significance \cite{Lipkin64}. However, it was later recognized as a measure characterizing chiral systems \cite{Tang10}. 
In comparison to the zilch of the EM field, the electron chirality density $\tau^Z$
can be regarded as the ``zilch of matter field'' \cite{Hoshino23}.

The three-component vectors in pseudospin representation are also constructed by the combination with the Pauli matrix in real-spin space: $\tau^0\bm \sg,\tau^X\bm \sg,\tau^Y\bm \sg,\tau^Z\bm \sg$, each of which corresponds to $\bm {\mathcal A},\bm {\mathcal M},\bm {\mathcal P},\bm {\mathcal V}$, respectively, to complete the Dirac bilinears given in Eqs.~(\ref{eq:DB_S}--\ref{eq:DB_T}).
The symmetries are intuitively identified with the pseudospin representation.
Namely, the spatial inversion (SI) exchanges the right- and left-handed particles, $\psi_{R,L} \to \psi_{L,R}$, as the index $R/L$ originates from the helicity of electron ($\simeq$ inner product of momentum and spin angular momentum).
The time reversal (TR) operation, which is antiunitary, changes the sign of the Pauli matrix, $\bm \sg \to - \bm \sg$, as it originates from the spin angular momentum.
These symmetries are summarized in Fig.~\ref{fig:summary}.

\subsubsection{Lorentz scalar (\,$\mathcal S$)}

For completeness, let us also consider the Lorentz scalar given by
\begin{align}
\mathcal S &= \Psi^\dg \beta \Psi = \phi^\dg \phi - \chi^\dg \chi
.
\end{align}
The Gordon decomposition for the scalar just reproduces the Dirac equations for $\Psi$ and $\bar \Psi$.
Namely, by taking $A=1$ in Eq.~\eqref{eq:gordon_decomp_gen}, we obtain
\begin{align}
     2 mc^2 {\bar \Psi} \Psi
=
\hbar \Psi^\dg \ldvec{(\imu D_t)} \Psi 
     -  c \Psi^\dg \bm \al \cdot \dvec{\bm \Pi} \Psi
     .
\end{align}
We note that this equation is corrected by the anomaly effect in quantum field theory \cite{Fujikawa_book,Peskin_book} (see also Sec. II\,F).

The time derivative of the Lorentz scalar is also calculated to yield
\begin{align}
    \frac{\partial \mathcal S}{\partial t}
    &= \frac{c}{\hbar } \bar \Psi (-\imu \bm \al) \cdot \dvec{\bm \Pi} \Psi 
    \label{eq:scalar_eom}
    .
\end{align}
The right-hand side, which is regarded as an inner product of electric polarization $-\imu \bm \al$ and momentum $\bm \Pi$, gives a source term of the Lorentz scalar.
This 
is analogous to the inner product of magnetization and momentum in Eq.~\eqref{eq:A0_Gordon}, which provides a helicity density.
We will briefly revisit Eq.~\eqref{eq:scalar_eom} in Sec.~V.

\subsection{Conjugate fields}

Let us identify the EM fields conjugate to Dirac bilinears.
Specifically, 
the external field coupled to the electron chirality $\tau^Z$ can `charge' the chirality, as will be utilized as an EM control of the chirality of materials.
For example, the chirality-induced phenomena such as MCA \cite{Tokura18} and CISS \cite{Gohler11} may be observed under the irradiation of light.

While the coupling between the electrons and EM fields is usually considered in NRL for condensed matter physics, we can also address this problem based on the Dirac equation without taking any limit.
We begin with the Dirac equation written in the following form:
\begin{align} 
\imu \gm^0 \partial_t \Psi  &= 
\Big[ \bm \gm \cdot (\bm p - e\bm A) + m + e\Phi \gm^0 \Big] \Psi
.
\end{align} 
In this subsection, we take the natural unit $\hbar=c=1$ for clarity.
This first-order differential equation clearly shows that the conjugate field to the charge ($\sim e\gm^0$) is the scalar potential $\Phi$, and the one conjugate to the current ($\sim e\bm \gm$) is the vector potential $\bm A$.
The mass term is also interpreted as conjugate to the Lorentz scalar ($\sim 1$).
Thus, the conjugate EM fields for a part of Dirac bilinears, Eqs.~\eqref{eq:DB_V} and \eqref{eq:DB_S}, are identified.

Next, we consider the conjugate fields for the other physical quantities.
However, this is not apparently seen in the above Dirac equation.
On the other hand, intuitively, we naively expect that the electric polarization and magnetization are conjugate quantities to electric and magnetic fields, respectively.
Actually, this property can be seen in the second-order equation:
\begin{align}
(\imu \gm^0 D_t)^2\Psi  &= 
\Big[ \bm \Pi^2 + m^2 
- e\bm \Sigma \cdot \bm B
- e(-\imu \bm \al) \cdot \bm E 
\Big] \Psi
,
\end{align}
which is, in the absence of EM fields, known as the Klein-Gordon equation \cite{Dirac_book}.
This second-order equation shows that the magnetization $\sim e\bm \Sigma$ and the electric polarization $\sim e(-\imu \bm \al)$ are indeed conjugate to the magnetic field $\bm B$ and electric field $\bm E$, respectively [see also Eq.~\eqref{eq:DB_T}].

In the second-order equation, the chirality and its conjugate field are still missing.
The coupling relevant to chirality appears in the third-order equation:
\begin{align}
(\imu \gm^0 D_t)^3 \Psi  &= 
m (\bm \Pi^2 + m^2 )\Psi + e\gm^0 (3\imu \bm E\cdot \bm \Pi + {\rm div\,} \bm E)\Psi 
\nonumber \\
&\hspace{-8mm}
+\bm \gm \cdot \qty[ \bm \Pi (\bm \Pi^2 + m^2 ) + e (-\dot {\bm E} - {\rm rot\,} \bm B + \imu \bm B\times \bm \Pi) ]\Psi
\nonumber \\
&\hspace{-8mm}
+ me \qty[ (-\imu\bm\al)\cdot \bm E - \bm \Sigma \cdot \bm B ] \Psi
\nonumber \\
&\hspace{-8mm}
+ e\gm^0 \gm^5 \bm B\cdot \bm \Pi \Psi
+ e\bm \gm\gm^5 \cdot (\bm E\times \bm \Pi 
+ \imu \dot {\bm B}
) \Psi
, \label{eq:3rd_Dirac}
\end{align}
where the dot ($\dot \ $) symbol represents the time derivative ($\dot O = \partial_t O$).
The first line shows the conjugate quantities of the scalar and charge.
The second line of Eq.\eqref{eq:3rd_Dirac} shows the quantities that couple to the current, and the third line includes the electric polarization and magnetization.

The interesting information can be obtained by looking at the fourth line of Eq.~\eqref{eq:3rd_Dirac}.
With $\bm \Pi = \bm p - e \bm A$ in mind, 
the conjugate field of the time component of pseudovector $\gm^0 \gm^5$, i.e. chirality density, is identified as $\bm A\cdot \bm B$,  which is a chirality of EM field and is known as magnetic helicity (density) \cite{Elsasser56,Woltjer58,Coles12}.
Also, the space component $\bm \gm \gm^5$, i.e. axial current, is conjugate to $\bm E\times \bm A$, which is known as the spin operator of photons \cite{Coles12,Gross_book}.
In a four component formalism, it is convenient for EM fields to define the helicity density and its related current by \cite{Ranada92}
\begin{align}
    \tilde A^{\mu} &= \frac 1 2 \epsilon^{\mu\nu\lambda \rho} A_\nu F_{\lambda \rho}
    = \frac 1 2 \tilde F^{\mu\nu} A_\nu
    , \label{eq:def_of_tilde_A}
\end{align}
which serves as the chiral vector potential conjugate to four-component pseudovector.
Here, $\epsilon$ is the totally antisymmetric tensor with $\epsilon^{0123}=+1$.
Indeed, we identify $\tilde A^\mu = (\bm A\cdot \bm B, \bm E \times \bm A+\Phi\bm B )$.
The equation of continuity is given by
 $\partial_\mu \tilde A^\mu = -2\bm E\cdot \bm B$ \cite{Berger84}, where the right-hand side has a same form as the chiral anomaly in Eq.~\eqref{eq:eom_chirality_full_relat} \cite{Akamatsu13}.

There is still one missing Dirac bilinear in the present context, which is the pseudoscalar $\mathcal P$.
Actually, if we consider the fourth-order equation,
the $\bm E\cdot \bm B$ term is identified as a conjugate quantity to the pseudoscalar.
To see this, instead of writing down the full fourth-order equation, we focus on a part of it as follows:
\begin{align}
    (\imu \gm^0 D_t)^4 \Psi &= \imu \gm^0 D_t(e\gm^0\gm^5 \bm B\cdot \bm \Pi \Psi) + (\text{others})
    \\
    &= e \imu \gm^5(e\bm B\cdot \bm E + \dot {\bm B} \cdot \bm \Pi)\Psi
    - e (\bm B\cdot \bm \Pi) (\bm \Sigma \cdot \bm \Pi) \Psi \nonumber \\
    &\ \  - me \gm^0 \gm^5 \bm B\cdot \bm \Pi \Psi 
    + (\text{others})
    ,
\end{align}
where the coupling between the pseudoscalar ($\imu \gm^5$)
and $\bm E\cdot \bm B$ ($=- F_{\mu\nu}\tilde F^{\mu\nu} / 8$) is clearly seen.
Thus, the higher-order Dirac equations are helpful to understand the conjugate fields.
In Sec.~III\,D, we will also discuss the conjugate fields based on the Hamiltonian with relativistic corrections by taking NRL.

\subsection{Reconsideration of Maxwell equation}

Below, we demonstrate that another microscopic physical quantities show up by reconsidering 
the Maxwell equation.
Originally, the Maxwell equation takes the following form:
\begin{align}
    \bm \nabla\cdot \bm E &= 4\pi \rho
    ,\label{eq:EB1}
    \\
    \bm \nabla\cdot \bm B &= 0
    ,\label{eq:EB2}
    \\
    \bm \nabla\times \bm E &= - \frac 1 c  \frac{\partial \bm B}{\partial t}
    ,\label{eq:EB3}
    \\
    \bm \nabla\times \bm B &= \frac{4\pi}{c} \bm j + \frac 1 c \frac{\partial \bm E}{\partial t}
    .\label{eq:EB4}
\end{align}
We call this the $\bm E$-$\bm B$ representation.
With the Gordon decomposition of charge and current in Eqs.~\eqref{eq:gordon_charge} and \eqref{eq:gordon_current}, it can be rewritten as \cite{Rodina08}
\begin{align}
    \bm \nabla\cdot \bm D &= 4\pi \rho_{\rm c}
    ,\label{eq:DH1}
    \\
    \bm \nabla\cdot \bm H &= 4\pi \rho_{\rm m}
    ,\label{eq:DH2}
    \\
    \bm \nabla\times \bm D &= \frac{4\pi}{c} \bm j_{\rm m}
    - \frac 1 c \frac{\partial \bm H}{\partial t}
    ,\label{eq:DH3}
    \\
    \bm \nabla\times \bm H &= \frac{4\pi}{c} \bm j_{\rm c} + \frac 1 c \frac{\partial \bm D}{\partial t}    
    ,\label{eq:DH4}
\end{align}
where we have defined
\begin{align}
    \bm D &= \bm E + 4\pi \bm P ,
    \\
    \bm H &= \bm B - 4\pi \bm M ,
\end{align}
which
are combined quantity of EM ($\bm E,\bm B$) and matter ($\bm P,\bm M$) fields.
These expressions provide the $\bm D$-$\bm H$ representation.
The electric charge density $\rho_{\rm c}$ and current density $\bm j_{\rm c}$ has already defined in Eqs.~\eqref{eq:convective_charge} and \eqref{eq:convective_current}, respectively, while
we have additionally defined
\begin{align}
\rho_{\rm m} &= - \bm \nabla \cdot \bm M ,
\\
\bm j_{\rm m} &=    c \bm \nabla \times \bm P 
    - \frac{\partial \bm M}{\partial t} ,
    \label{eq:jm_orig_def}
\end{align}
which are magnetic charge and current densities, respectively.
The parity signs are $({\rm SI/TR})=(+/+)$ for $\rho_{\rm c},\bm j_{\rm m}$ and $({\rm SI/TR})=(-/-)$ for $\rho_{\rm m},\bm j_{\rm c}$.
The fields $\bm D$ and $\bm H$ are usually introduced as the electric flux density and magnetic field for a macroscopic equation.
In the antenna theory, the magnetic current is defined as a convenient object to describe macroscopic EM fields \cite{Stratton_book}.
Here, however, these quantities are {\it microscopically} defined in terms of EM and Dirac fields.

The magnetic current is further rewritten by using the equation of motion for the magnetization:
\begin{align}
    \bm j_{\rm m} &= - \frac{e}{2m} \bar \Psi \dvec{\bm \Pi} (\imu \gm^5) \Psi 
    . \label{eq:def_of_jm_Dirac}
\end{align}
This expression is a combination of the Lorentz pseudoscalar $\imu\gm^5$ and the canonical momentum $\bm \Pi$, and may be interpreted as a flow of pseudoscalar.
A similar interpretation can be applied to $\bm j_{\rm c}$ in Eq.~\eqref{eq:convective_current}, which is a flow of Lorentz scalar.
In a similar manner, a complementary representation for the magnetic charge can also be constructed as follows:
\begin{align}
    \rho_{\rm m} &= - \frac{\hbar e}{2mc^2} \bar \Psi (\ \ldvec{\imu D_t}\ )\imu \gm^5 \Psi
    . \label{eq:def_of_rhom_Dirac}
\end{align}
Equations \eqref{eq:def_of_rhom_Dirac}
 and \eqref{eq:def_of_jm_Dirac} are compared to Eqs.~\eqref{eq:convective_charge} and \eqref{eq:convective_current}, where the Lorentz scalar ($\bar \Psi \Psi$) is replaced by the pseudoscalar ($\bar \Psi \imu \gm^5 \Psi$).

Let us also introduce the four component vector representation \cite{Gordon28}:
\begin{align}
    \partial_\nu G^{\mu\nu} &= - \frac{4\pi}{c} j^\mu_{\rm c}
    , \label{eq:G-field1}
    \\
    \partial_\nu \tilde G^{\mu\nu} &= - \frac{4\pi }{c}j^\mu_{\rm m}
    , \label{eq:G-field2}
\end{align}
where the antisymmetric tensors are defined by
\begin{align}
    G^{\mu\nu} &= F^{\mu\nu} - 4\pi T^{\mu\nu}
    , \\
    \tilde G^{\mu\nu} &= \epsilon^{\mu\nu\lambda \rho} G_{\lambda \rho}
\end{align}
with $j_{\rm c,m}^\mu = (c \rho_{\rm c,m},\bm j_{\rm c,m})$, $T^{\mu\nu} = - \frac{\hbar e}{2mc} \mathcal T^{\mu\nu} = (\bm P,\bm M)$ and $j_{\rm m}^\mu = c \partial_\nu \tilde T^{\mu\nu}$.
Since $G$ and $\tilde G$ are antisymmetric tensors, we have the equation of continuity $\partial_\mu j^\mu_{\rm c} = \partial_\mu j_{\rm m}^\mu = 0$ as derived from Eqs.~\eqref{eq:G-field1} and \eqref{eq:G-field2}.
Namely, in addition to the electric charge, the magnetic charge is also a conserved quantity in general.
The above relations may be contrasted against the usual Maxwell equations $\partial_\nu F^{\mu\nu} = - \frac{4\pi}{c}j^\mu$, $\partial_\nu \tilde F^{\mu\nu} = 0$, and $\partial_\mu j^\mu =0$.

\subsection{Correction by quantum field theory: Quantum anomaly}

If we consider the quantum field theory, 
the path integral is needed for the expectation value of physical quantities, which corresponds to the quantum mechanical average over many body states.
In this case,
a correction is needed for the two of the relations in the last subsections \cite{Fujikawa_book}.
More specifically, the Gordon decomposition of the scalar ($\mathcal S$) and pseudoscalar ($\mathcal P$) is corrected in the presence of EM fields.
The derivation using path-integral is given in Ref.~\cite{Fujikawa_book}, and Appendix A provides a derivation in terms of statistical mechanics.

We have the modified equation of continuity for chirality density as
\begin{align}
    &\hbar \frac{\partial \la \mathcal A^0 \ra}{\partial t} + \hbar c \bm \nabla \cdot \la \bm {\mathcal A}\ra = 2mc^2 \la \mathcal P\ra - \frac{e^2}{2\pi^2\hbar c} \bm E\cdot\bm B
    , \label{eq:eom_chirality_full_relat}
\end{align}
where the bracket means the quantum statistical average of a many-body state.
The final term stems from the chiral anomaly.
Another correction is known as the Weyl or trace anomaly:
\begin{align}
& \hbar \la \Psi^\dg \ldvec{(\imu D_t)} \Psi \ra
     -  c \la \Psi^\dg \bm \al \cdot \dvec{\bm \Pi} \Psi\ra
     =
     2 mc^2 
     \la \mathcal S \ra
     - \frac{e^2}{6\pi^2\hbar c} (\bm E^2 - \bm B^2)    
     . \label{eq:eom_scalar_full_relat_anomaly}
\end{align}
We will derive these relations for NRL in Sec. III\,F.

We note that, if the Dirac equation were naively used, the terms other than the EM field parts (i.e., $\bm E\cdot \bm B$ and $\bm E^2-\bm B^2$)
in the above relations would vanish.
In order to avoid this contradiction, one needs to consider the difference of the coordinates for $\bar \Psi(x)$ and $\Psi(x')$ with $x = (ct,\bm r)$, where their product has a singularity in the limit $x\to x'$ \cite{Peskin_book,Fujikawa_book}.

\section{Non-relativistic limit and Relativistic corrections}

\subsection{Hamiltonian }

In order to discuss the relativistic correction for the Dirac bilinears and other microscopic quantities, it is convenient to derive the explicit relation between four-component Dirac field and two-component Schr\"odinger field in NRL.
While the effective Hamiltonian in NRL is derived based on Foldy-Wouthuysen (FW) transformation \cite{Foldy50,Tani49, Silenko16},
here we show the derivation by employing the Berestetskii-Landau method \cite{Berestetskii49,Berestetskii_book}, which is a straightforward formalism to obtain the two-component equation (so-called elimination method \cite{Silenko08}).
In Ref.~\cite{Hoshino23}, we dealt with the $1/c$ expansion, but in the following we derive the expressions by $1/m $ expansion where the gauge invariance is manifestly present.

We consider the Dirac equation given in Eq.~\eqref{eq:Dirac_Equation}.
The differential equation for $\chi$ defined by Eq.~\eqref{eq:def_of_phi_chi} can be formally solved as
\begin{align}
    \chi(t) &= \frac{c}{\imu \hbar} \int^t \diff t' \, \epn^{ 2\imu mc^2  (t-t')/\hbar } 
    \nonumber \\
    &\hspace{10mm} \times
    \exp[ \frac{\imu e}{\hbar} \int_t^{t'} \diff t'' \Phi(t'') ] X(t') \phi(t')
    , \label{eq:chi_general}
\end{align}
where the time-dependence is explicitly shown.
We have introduced the shorthand notation $X = \bm \sg \cdot \bm \Pi$.
In NRL with $m\to\infty$ or $c\to \infty$, the phase factor is rapidly oscillating to vanish.
We expand the expression by assuming that the rest-mass energy $mc^2$ is large, which is performed by repeating the integration by parts in Eq.~\eqref{eq:chi_general}, and obtain
\begin{align}
    \chi &= \frac{1}{2mc} \sum_{n=0}^\infty \qty( \frac{\hbar}{2\imu mc^2} )^n D_t^n X \phi
    . \label{eq:chi_expand}
\end{align}
This result is also obtained by a formal expansion of the operator $(\imu \hbar D_t + 2mc^2)^{-1}$.
The above solution is inserted into the equation for $\phi$, resulting in
\begin{align}
    \imu \hbar D_t \phi = \frac{1}{2m} \sum_{n=0}^\infty \qty(\frac{\hbar}{2\imu mc^2})^n X D_t^n X \phi
    .\label{eq:Sch1}
\end{align}
The wave function in the Schr\"odinger equation must satisfy the norm conservation condition \cite{Berestetskii_book}:
\begin{align}
    \int \diff \bm r \Psi^\dg \Psi 
    = \int \diff \bm r \qty( \phi^\dg \phi + \chi^\dg \chi)
    = \int \diff \bm r \psi^\dg \psi
    ,
\end{align}
where 
$\psi$ is a norm-conserved two-component Schr\"odinger field.
Based on this condition, the relation between $\phi$ and $\psi$ can be obtained up to the third order of $1/m$ as the sufficient condition
\begin{align}
    \psi = \qty(1  + \frac{X^2}{8m^2c^2} + \frac{\hbar XD_t X}{8\imu m^3c^4}) \phi + O(m^{-4})
    . \label{eq:phi_to_psi}
\end{align}
By substituting this result into Eq.~\eqref{eq:Sch1}, we obtain the Schr\"odinder equation.
Defining $Y= [D_t, X]$, the equation is given by 
\begin{align}
\hspace{-2mm}
    \imu \hbar D_t \psi = \qty( \frac{X^2}{2m} + \frac{\imu \hbar [Y,X]}{8m^2 c^2} - \frac{X^4}{8m^3c^2} + \frac{\hbar^2 Y^2}{8m^3c^4} ) \psi
    + O(m^{-4})
    . \label{eq:effective_hamiltonian_up_to_order_3}
\end{align}
The explicit form can be evaluated by the concrete forms of $X^2$ and $Y$:
\begin{align}
    X^2 &= \bm \Pi^2 - \frac{\hbar e}{c} \bm \sg \cdot \bm B
    , \\
    Y &= e\bm \sg \cdot \bm E
    .
\end{align}
Thus, the expectation value of the energy, or the Hamiltonian operator, is written as
 \begin{align}
\mathscr H &=  \mathscr H^{(0)} +  \mathscr H^{(1)} +  \mathscr H^{(2)} + \mathscr H^{(3)} + O(m^{-4})
,
 \end{align}
where the zeroth-order term is given by
 \begin{align}
 \mathscr H^{(0)} &= \int \diff \bm r \psi^\dg e\Phi \psi = \int \diff \bm r \rho^{(0)} \Phi
 . \label{eq:ham0}
 \end{align}
 This shows a coupling between charge and scalar potential.
 The first-order Hamiltonian is
 \begin{align}
 \mathscr H^{(1)} &= \int \diff \bm r \psi^\dg \qty( 
 \frac{1}{2m} \bm \Pi^2 - \frac{\hbar e}{2mc} \bm B\cdot \bm \sg  ) \psi
 ,\label{eq:ham1}
 \end{align}
which is composed of the usual kinetic energy and the Zeeman term.
The second-order contributions are
\begin{align}
    \mathscr H^{(2)}
    &= \int \diff \bm r \psi^\dg \bigg( 
- \frac{\hbar^2 e}{8m^2c^2} {\rm div\,} \bm E
\nonumber \\
&\ \ \ 
    -\frac{\hbar e}{8m^2c^2} \big[ 
\bm\Pi \cdot (\bm \sg \times \bm E) + (\bm \sg \times \bm E) \cdot \bm\Pi \big] 
\bigg) \psi
 . \label{eq:ham2}
\end{align}
The first term corresponds to the Darwin term, which is interpreted as a quantum mechanical correction to the scalar potential \cite{Baym_book}.
The second line is the spin-orbit coupling, which generates a number of interesting phenomena in condensed matter physics.

Finally, the third-order contributions are
\begin{align}
    \mathscr H^{(3)} &= \int \diff \bm r \psi^\dg \bigg(
    - \frac{\bm \Pi^4}{8m^3c^2}
    + \frac{\hbar^2 e^2}{8m^3c^4}(\bm E^2- \bm B^2)
    \nonumber \\
    &\ \ \ 
+\frac{\hbar e}{8m^3c^3} \big[ \bm \Pi^2 (\bm \sg \cdot \bm B) + (\bm \sg \cdot \bm B) \bm \Pi^2 \big]
    \bigg)\psi
 , \label{eq:ham3}
\end{align}
where the first term is a correction to the kinetic energy, and the second term is a correction to the scalar potential.
The third term is regarded as a correction to the Zeeman term.

The above expression has already been obtained in the previous studies \cite{Frohlich93}.
For our purpose, it is also necessary to evaluate the physical quantities such Dirac bilinears.
Hence, we show the expression for the $\phi$ and $\chi$ fields, which are explicitly calculated based on Eqs.~\eqref{eq:chi_expand}, \eqref{eq:Sch1} and \eqref{eq:phi_to_psi}:
\begin{align}
    \phi &= \qty(1 - \frac{X^2}{8m^2c^2} - \frac{\hbar XY}{8\imu m^3c^4})\psi + O(m^{-4})
    ,\\
    \chi &= \qty(\frac{X}{2mc} + \frac{\hbar Y }{4\imu m^2c^3}
    - \frac{3X^3}{16m^3c^3} - \frac{\hbar^2 \dot Y}{8m^3c^5} )\psi + O(m^{-4})
    ,
\end{align}
where $\dot Y =\partial_t Y = e\bm \sg \cdot \dot {\bm E}$.

\subsection{Electric charge and current}

Let us first derive the leading-order contributions for $\bm M$ and $\bm P$ in NRL, which are related to the charge and current as shown in Eqs.~\eqref{eq:gordon_charge} and \eqref{eq:gordon_current}. The straight-forward calculation gives \cite{Wang06}
\begin{align}
    \bm M &\simeq \bm M^{(1)} = \frac{\hbar e}{2mc} \psi^\dg \bm \sg \psi \equiv \bm M_S
    ,\\
    \bm P &\simeq \bm P^{(2)} = 2\qty[ \bm P_S  - \frac{\hbar^2e}{8m^2c^2} \bm \nabla (\psi^\dg \psi) ]
    .
\end{align}
Throughout this paper, the superscript number in parenthesis, which is put on the physical quantities defined in terms of Dirac four-component fields, represents the order of $m^{-1}$ for relativistic corrections.
The spin-derived electric polarization is defined by
\begin{align}
    P_{Si} &\equiv 
    \frac{\hbar e}{8m^2c^2} \psi^\dg (\ \dvec{\bm \Pi}\times \bm \sg)_i \psi
    =
    \frac{\hbar e}{8m^2c^2} \epsilon_{ijk} j_{Sjk}
    ,\\
    j_{Sij} &\equiv \psi^\dg  \dvec{\Pi}_i \sg^j \psi
    ,
\end{align}
where $i,j,k = x,y,z$.
$j_{Sij}$ is the spin-current tensor which shows a flow in $i$-direction of spin in $j$-direction.
Note that the subleading order contribution for $\bm M$ is $O(m^{-3})$.

The spin current tensor may be decomposed into scalar, vector (or antisymmetric tensor), and symmetric tensor as \cite{Hayami18}
\begin{align}
    j_{Sij} &= \frac{\delta_{ij}}{3} \psi^\dg \dvec{\bm \Pi}\cdot \bm \sg \psi
    + \frac{\epsilon_{ijk}}{2} \psi^\dg (
    \  \dvec{\bm \Pi}\times \bm \sg)_k \psi
    + j_{Sij}^{\rm sym}
    ,\label{eq:spin_curr_decomposition}
\end{align}
where $J_{Sij}^{\rm sym}$ is a symmetric and traceless matrix.
The first term is identified as a helicity in NRL, and the second term is a spin-derived electric polarization.
The third term may be interpreted as describing anisotropy of the helicity.

Next, let us consider the charge and current densities.
Taking NRL of charge density, we obtain
\begin{align}
    &\rho^{(0)} = e \psi^\dg \psi = e \mathcal S^{(0)}
    ,\label{eq:S_NRL}
\end{align}
which is identical to the Lorentz scalar in NRL.
The next leading-order contribution is given by
\begin{align}
    \rho^{(2)} &= 
    \rho_{\rm c}^{(2)} - \bm \nabla \cdot \bm P^{(2)}
    =
    - \frac{1}{2} \bm \nabla \cdot \bm P^{(2)}
    . \label{eq:charge_second}
\end{align}
The charge density has another representation for the relativistic correction:
\begin{align}
    \rho &= \qty(1 + \frac{\hbar^2}{8m^2c^2} \bm \nabla^2 )\rho^{(0)} - \bm \nabla \cdot \bm P_S +O(m^{-3})
    . \label{eq:charge_another_form}
\end{align}
Namely, the gradient term of polarization is included into the charge density part.
In this case, the polarization is contributed only by the spin-derived quantity $\bm P_S$.
The correction in the first term of Eq.~\eqref{eq:charge_another_form} is physically interpreted as an uncertainty of position associated with the minimal length scale $\lambdabar = \hbar /mc$ (reduced Compton length) in relativistic quantum mechanics \cite{Baym_book}.
More specifically, the charge density is written as $\rho(\bm r + \bm \xi)$, where $\bm \xi$ is a positional fluctuation satisfying $\la \xi_i\ra = 0$ and $\la \xi_i\xi_j \ra = \lambdabar^2 \delta_{ij}/4$, and the $1/m$ expansion reproduces the first term of Eq.~\eqref{eq:charge_another_form}.

The current density is written as $\bm j = \bm j^{(1)} + \bm j^{(2)} + \bm j^{(3)}  + O(m^{-4})$.
The first-order term in NRL is given by
\begin{align}
    \bm j^{(1)} &= 
    \frac{e}{2m} \qty[ \psi^\dg \bm \sg (\bm \sg\cdot \bm \Pi)\psi + {\rm conj.} ]
    ,\label{eq:j1-1}
    \\
    &=\frac{e}{2m} \psi^\dg \dvec{\bm \Pi} \psi
    + c\bm \nabla \times \bm M_S
    ,\label{eq:j1-2}
\end{align}
where the intermediate expression [$\sim \bm \sg (\bm \sg \cdot \bm \Pi)$] is written for reference.
Note that the current expression is different for $1/m$ expansion and $1/c$ expansion; the latter case does not capture the vector potential $\bm A$ contribution included in $\bm \Pi$.
In response to a uniform magnetic field, the circular current is produced to generate diamagnetism in the first term and paramagnetism in the second term, both of which contributes to the experimentally observed magnetization \cite{Fukuyama71,Ozaki21}.

The second-order term is given by
\begin{align}
    \bm j^{(2)} &= \frac{e}{mc} \bm E \times \bm M_S
    .\label{eq:j2}
\end{align}
This is a Hall-current like contribution which becomes finite in the presence of magnetization.
The third-order contribution
is given by
\begin{align}
    \bm j^{(3)}
     &= \frac{\hbar e^2}{2m^2c^2} H \bm B
     + \frac{e}{2mc} \bm P^{(2)} \times \bm B
     - \frac{\hbar^2 e}{4m^3c^4}\rho^{(0)} \dot {\bm E}
     \nonumber \\
     &\hspace{-2mm}
     - \frac{e}{16m^3c^2} \qty[ 3\psi^\dg \bm \Pi^2 + \qty( \bm \Pi^{*2}\psi^\dg ) ] \bm \sg (\bm \sg \cdot \bm \Pi) \psi + {\rm conj.}
     ,\label{eq:j3}
\end{align}
where we have defined the helicity density in NRL:
\begin{align}
    H &\equiv  \frac{1}{2mc} j_{Sii} = \frac{1 }{2mc} \psi^\dg \dvec{\bm \Pi} \cdot \bm \sg \psi
    ,\label{eq:def_of_helicity_NRL}
\end{align}
which is a dimension of the inverse of volume.
The second line of Eq.~\eqref{eq:j3} is regarded as a correction to $\bm j^{(1)}$ as is obvious from the structure $\bm \sg (\bm \sg \cdot \bm \Pi)$ which appears also in Eq.~\eqref{eq:j1-1}.
On the other hand, the first line of Eq.~\eqref{eq:j3} is a qualitatively new contribution to the current, which is directly proportional to the magnetic field.
With this term, at first sight, the uniform magnetic field could act as a voltage to produce the electric current. 
In this regard, the first term proportional to $H \bm B$ is an `Ohm' contribution and the second one with $\bm P^{(2)}\times \bm B$ is a `Hall' contribution.

The induction of electric current by magnetic field is known as a chiral magnetic effect \cite{Fukushima08}.
However, such current cannot be finite in equilibrium \cite{Zhou13,Vazifeh13}. 
The absence of the current is also known as a consequence of the Bloch-Bohm theorem for both NRL \cite{Bohm49,Ohashi96,Yamamoto15,Watanabe19} and relativistic case \cite{Yamamoto15}.
Hence, the contributions which counteract against  $H \bm B$ and $\bm P^{(2)}\times \bm B$ are necessarily present.
We will revisit the Bloch-Bohm theorem in Sec.~IV.
We note that the magnetic-field-driven current can exist in non-equilibrium situations \cite{Sekine21}.

In relation to the equation of continuity, let us consider the time-derivative of charge density and polarization up to $O(m^{-2})$:
\begin{align}
    \frac{\partial \rho^{(0)}}{\partial t}
    &= -\bm \nabla \cdot \bm j^{(1)} - \frac{1}{2} \bm \nabla \cdot \bm j^{(2)} + O(m^{-3})
    ,\label{eq:eoc_NRL}
    \\
    \frac{\partial \bm P^{(2)}}{\partial t} &= \bm j^{(2)} + O(m^{-3})
    .
\end{align}
Combining these with Eq.~\eqref{eq:charge_second}, we identify the equation of continuity
\begin{align}
    \frac{\partial \qty( \rho^{(0)} + \rho^{(2)})}{\partial t}
    +\bm \nabla \cdot \qty( \bm j^{(1)} + \bm j^{(2)} ) =  O(m^{-3})
    , \label{eq:eoc_charge}
\end{align}
which is consistent with the perturbative expansion for the equation of continuity of $j^\mu = (c\rho,\bm j)$ given in Eq.~\eqref{eq:eoc_charge_Dirac}.

\subsection{Chirality density and axial current}

Let us derive the relativistic corrections for chirality and axial current densities defined in Eqs.~\eqref{eq:Dirac_chirality} and \eqref{eq:Dirac_chirality_current}.
Performing $1/m$ expansion, we obtain
\begin{align}
    \tau^Z = -\mathcal A^0 &\simeq \tau^{Z(1)} = H
    .
\end{align}
Namely, the chirality density inherent in the Dirac equation is identical to the helicity density $H$ defined in Eq.~\eqref{eq:def_of_helicity_NRL}.
While it has been well-known that the chirality is identical to the helicity in massless limit ($m\to 0$) \cite{Gross_book},
the relation between $\gm^5$ and helicity holds even in the expansion from NRL \cite{Hoshino23}.
Note that the next leading-order contribution for the electron chirality density is $O(m^{-3})$.

The relativistic corrections for other chirality-related quantities are given by
\begin{align}
    &\bm {\mathcal A} \simeq \bm {\mathcal A}^{(0)} = - \psi^\dg \bm \sg \psi
    = - \frac{2mc}{\hbar e} \bm M_S
    ,\label{eq:A_NRL}
    \\
    & \mathcal P \simeq \mathcal P^{(1)} = - \frac{1}{e} \bm \nabla \cdot \bm M_S
    .\label{eq:P_NRL}
\end{align}
Namely, the axial current $\bm {\mathcal A}$ is identified as the spin density in the leading-order approximation.
The pseudoscalar is identified as the divergence of magnetization.

For the equation of continuity of chirality, it is necessary to keep the higher-order term, since the leading-order $O(1)$ contributions exactly cancel to each other \cite{Banerjee20}, as confirmed from Eqs.~\eqref{eq:A_NRL} and \eqref{eq:P_NRL}.
We keep the contribution up to $O(m^{-2})$, and obtain
\begin{align}
    \frac{\partial \tau^{Z(1)} }{\partial t} + \bm \nabla \cdot \bm j^Z = \frac{2}{\hbar} \bm E\cdot \bm M_S
    + O(m^{-3})
    ,\label{eq:eom_NRL}
\end{align}
which is a chirality version of Eq.~\eqref{eq:eoc_NRL}.
We have defined the chirality current density
\begin{align}
    \bm j^Z &\equiv \frac{1}{4m^2c} \qty( \psi^\dg \dvec{\bm \Pi} \big( \dvec{\bm \Pi}\cdot \bm \sg \big) \psi + \frac \hbar 2 \bm \nabla \times \psi^\dg \dvec{\bm \Pi} \psi )
    ,\label{eq:jz-def}
\end{align}
which is not identical to the NRL expression of the axial current, although they are related.
For the derivation, we have used $\tau^{Z(2)} = 0$ and $\bm {\mathcal A}^{(1)} = 0$, and 
the quantities $\tau^{Z(1)}, \bm {\mathcal A}^{(2)}, \mathcal P^{(2,3)}$ are involved in Eq.~\eqref{eq:eom_NRL}.
It is also notable that the first term of Eq.~\eqref{eq:jz-def} is a leading-order relativistic correction for $\Psi^\dg \dvec{\bm \Pi} \gm^5 \Psi$, which is regarded as a flow of the chirality density.
The rotation part of Eq.~\eqref{eq:jz-def}
, which is 
is not apparently dependent on the spin, derives from $\bm {\mathcal A}^{(2)}$.
This is analogous to the vorticity $\bm \nabla \times \bm v$ in fluid mechanics with $\bm v$ being the velocity field, which contributes to the chirality current according to Eq.~\eqref{eq:jz-def}.
The vorticity part does not contribute to the equation of continuity as is also the case for $\bm \nabla \times \bm M_S$ in the usual current [see Eq.~\eqref{eq:j1-2}].

Next, we consider the chirality polarization defined in Eq.~\eqref{eq:def_chiral_polarization}.
The leading-order relativistic corrections are given by
\begin{align}
    \bm {\mathcal P}_C &\simeq \bm {\mathcal P}_C^{(1)}
    = \frac{\hbar c}{2} \bm \nabla H - \frac 1 {4m} 
    \psi^\dg \ldvec{(\imu \bm \Pi^2)} \bm \sg \psi
     + \bm M_S\times \bm B
     .
\end{align}
In the time-reversal symmetric system, the last term in the right-hand side with magnetic field $\bm B$ vanishes, and then the spatial integration is zero in general, $\int \diff \bm r \bm {\mathcal P}_C = 0$, unless the boundary of system is considered. [see also Eq.~\eqref{eq:Axial_cur_t_deriv}].

The electron chirality $\tau^{Z(1)} = H$
involve the spin degrees of freedom ($\bm \sg$).
Actually, we can also employ the definition of chirality without involving spin and EM fields if we consider the two-body quantities.
Namely, we consider the hydrodynamical helicity \cite{Moreau61,Moffatt69}, which is defined by the inner product of the velocity ($\sim \bm v$) and vorticity ($\sim \bm \nabla \times \bm v$).
We define the following quantity for electrons:
\begin{align}
    \mathcal H_{\rm hydro} &= \int \diff \bm r \, : \bm j\cdot (\bm \nabla \times \bm j) :
    \label{eq:hydro_helicity}
\end{align}
When the spinor field is promoted to an operator in the second-quantized formalism, it is natural to consider the normal ordering [denoted by colon (:) symbol] where the creation (annihilation) operators are put on the left (right).
Classically, Equation~\eqref{eq:hydro_helicity} is a quantity to measure the degree of linkage of the vortex lines \cite{Moffatt69}.

When we write the NRL current expression using the first term of Eq.~\eqref{eq:j1-2} [or, using $\bm j_{\rm c}$ in Eq.~\eqref{eq:jc_NRL_expression} instead of $\bm j$], we identify the chirality (helicity) which is not explicitly dependent on spin.
The concrete expression in NRL is obtained as 
\begin{align}
\mathcal H_{\rm hydro}^{(0)} &\sim
- \frac{\hbar^2e^2}{m^2}
\int \diff \bm r \,
\epsilon_{ijk} 
:(\psi^\dg \partial_i \psi) (\partial_j \psi^\dg \partial_k \psi) :
\label{eq:hydro_helic_NRL}
\end{align}
where only the Pauli-matrix-independent part is kept in the absence of magnetic field.
This expression can be finite even without spin degrees of freedom. Namely, even when we regard $\psi$ as a one-component (spinless) fermionic field, Eq.~\eqref{eq:hydro_helic_NRL} can be finite. 
On the other hand, if we considered a bosonic or classical scalar field, $\epsilon_{ijk}\partial_i \psi\partial_k\psi$ would vanish in general.
Hence, the fermionic nature is important for the spinless contribution to the hydrodynamic helicity.

The above quantity can also be used for the characterization of the chirality microscopically by looking at $\bm j\cdot (\bm \nabla \times \bm j)$ at every spatiotemporal point.
Furthermore, the other helicity-like quantities may also be introduced by using the three-component vectors as, e.g. $\bm M\cdot (\bm \nabla \times \bm M)$ and/or $\bm P\cdot (\bm \nabla \times \bm P)$, which is analogous to the Lipkin's zilch $C = \bm B\cdot (\bm \nabla \times \bm B) + \bm E\cdot (\bm \nabla \times \bm E)$ \cite{Lipkin64,Tang10}.
In these examples, the spin degrees of freedom are explicitly involved in the leading-order relativistic correction as distinct from $\mathcal H_{\rm hydro}$.

In Appendix B, we provide a simple example of chiral/polar systems, which shows finite values of $\la H\ra$ and $\la \bm P_S \ra$.
We have considered the uniform system with spin-orbit coupling for electron gas in equilibrium, where the current $\la \bm j\ra$ and magnetization $\la \bm M_S \ra$ are zero due to the time-reversal symmetry.
Nevertheless, we obtain the finite value of $\mathcal H_{\rm hydro}$.
Although $\mathcal H_{\rm hydro}$ classically quantifies the number of knots of vortex lines, the present contribution enters only through the quantum mechanical effect.
The more detailed insight may be obtained by the path-integral formalism, which visualizes the quantum mechanical superpositions and connects the expression to the classical limit.
We emphasize that the definition of $\mathcal H_{\rm hydro}$ can be used also for non-uniform systems such as chiral molecules and crystals, which may quantify the chirality of the systems even without spin-orbit coupling.
It is also interesting to consider $\mathcal H_{\rm hydro}$ for nuclei, whose momentum is described by creation and annihilation operators of phonons \cite{Kittel_book}.

\subsection{Coupling to EM fields}

\subsubsection{Conjugate external fields}

In this section, we consider the conjugate EM fields in terms of $1/m$ expansion, which are identified in the relativistic corrections for the effective Hamiltonian.
In the linear response against the external magnetic field, we consider the perturbation in the form
\begin{align}
    \mathscr H_{\bm A} &= - \frac 1 c \int \diff \bm r \, \tilde {\bm j} \cdot \bm A + O(\bm A^2)
    , \label{eq:def_of_tilde_current}
\end{align}
which defines the current density $\tilde {\bm j}$ conjugate to the vector potential.
Here we assume a static vector potential $\bm A(\bm r)$ to describe the response to magnetic field.
The explicit form of the expression is written as
\begin{align}
    \tilde {\bm j}^{(1)} &= \frac{e}{2m} \psi^\dg \dvec{\bm \Pi} \psi
    + c\bm \nabla \times \bm M_S= \bm j^{(1)}
    ,
    \\
    \tilde {\bm j}^{(2)} &=  \frac{e}{2mc} \bm E \times \bm M_S = \frac 1 2 \bm j^{(2)}
    .
\end{align}
The NRL expression is identical to the electronic current itself: $\tilde {\bm j}^{(1)} = \bm j^{(1)}$.
On the other hand, notably, $\tilde {\bm j}$ is different from $\bm j$ in $O(m^{-2})$ terms ($\tilde {\bm j}^{(2)} \neq \bm j^{(2)}$).
We will discuss this discrepancy between $\bm j$ and $\tilde {\bm j}$ in more detail in Sec.~IV.

In addition, we can also consider the conjugate quantity to scalar potential:
\begin{align}
    \mathscr H_{\Phi} &= \int \diff \bm r \, \tilde \rho \Phi + O(\Phi^2)
    .
\end{align}
Its NRL expression is same as the charge density defined by the Dirac field up to $O(m^{-2})$, i.e., $\tilde \rho^{(0)} = \rho^{(0)}$ and $\tilde \rho^{(2)} = \rho^{(2)}$.
The conjugate quantities of $\bm M_S $ and $\bm P_S$ are, respectively, $\bm B$ and $\bm E$, which is obvious from Eqs.~\eqref{eq:ham1} and \eqref{eq:ham2} \cite{Hoshino23}.

Next, we search for the conjugate fields to electron chirality and the axial current, which is to be identified as 
the coupling to non-linear EM fields according to the discussion in Sec.~II\,D.
First, we write down the explicit form of the spin-orbit coupling in Eq.~\eqref{eq:ham2} and the higher-order Zeeman term in Eq.~\eqref{eq:ham3}:
\begin{align}
    \mathscr H'
    &=  \frac{\hbar e}{8m^2c^2} \int \diff \bm r \psi^\dg \bigg( 
    -
\bm\sg \cdot (\bm E \times \bm \Pi)
+\frac{1}{mc} (\bm \sg \cdot \bm B) \bm \Pi^2
\bigg) \psi
\nonumber \\
&\ \ \ 
+{\rm conj.}
 \label{eq:ham_chiral}
\end{align}
Noting that the NRL expression of axial current $\bm {\mathcal A}$ is identical to the spin density $\bm {\mathcal A}^{(0)} = - \psi^\dg \bm \sg \psi$ [Eq.~\eqref{eq:A_NRL}], the coupling to the axial current $\bm {\mathcal A}^{(0)}$ is the non-linear vector field $\bm E\times \bm A$.
This interpretation is also supported by the discussion in 
Sec. II\,D, where the spatial part of the axial current is coupled to its conjugate EM field  $\bm E\times \bm A$ without taking NRL [see also Eq.~\eqref{eq:3rd_Dirac}].

On the other hand, the second term of Eq.~\eqref{eq:ham_chiral} has a term proportional to $(\bm \sg \cdot \bm B) (\bm A\cdot \bm p)$, which is identical to 
$(\bm A\times \bm \sg) \cdot (\bm B\times \bm p) + (\bm A\cdot \bm B) (\bm p\cdot \bm \sg)$.
Thus, we identify the conjugate field to the chirality density $\tau^{Z(1)}=H\sim \bm p\cdot \bm \sg$ is the magnetic helicity density $\bm A\cdot \bm B$.
Another term with $(\bm A\times \bm \sg) \cdot (\bm B\times \bm p)$ includes the Lorentz-force like contribution $\bm v \times \bm B$ with $\bm v = \bm p/m$ (not identical to $\bm v$ defined in Sec.~II), which is coupled to $\bm A\times \bm \sg$.
Since the product of force and displacement gives energy (work),
we interpret $\bm A\times \bm \sg$ as a displacement vector, and such interpretation is indeed discussed in Refs.~\cite{Zou20,Ado24}.
See also Sec.~IV\,B for further discussions on position operator.

With these insights, we can rewrite the Hamiltonian as
\begin{align}
    &\mathscr H'_{(2)} = - 
    \frac{\hbar e^2}{4m^2c^3} 
    \int \diff \bm r \bigg(
    \bm {\mathcal A}^{(0)} \cdot (\bm E\times \bm A ) 
    - 2 H (\bm A\cdot \bm B)
    \nonumber \\
    &
    \hspace{20mm}
    - \frac{1}{mc}
    \psi^\dg (\bm A\times \bm \sg) \cdot (\bm B\times \bm p) \psi
    + \rm conj.
     \bigg)
     ,
\end{align}
where only the second-order terms with respect to fields are kept as indicated by the subscript (2).
The first line shows that the coupling constant $\sim \frac{\hbar e^2}{m^2c^3}$ is shared for both the spatial part $\bm {\mathcal A}^{(0)}$ and the time component $\mathcal A^{0(1)} = -H$ of the four-component pseudovector $\mathcal A^\mu$.

Let us also comment on the conjugate field to the Lorentz scalar in the view point of relativistic correction.
The third-order Hamiltonian given in Eq.~\eqref{eq:ham3} includes the following term:
\begin{align}
    \mathscr H'' &= \frac{\hbar^2 e^2}{8m^3c^4} \int \diff \bm r
    (\bm E^2- \bm B^2) \psi^\dg \psi
    . \label{eq:NRL_specific_2}
\end{align}
Since $\bm E^2-\bm B^2$ ($=- F_{\mu\nu}F^{\mu\nu}/4$) appears in the Lagrangian \eqref{eq:gen_Lagrangian}, which is a Lorentz scalar, the coupling in the density operator $\psi^\dg \psi$ in Eq.~\eqref{eq:NRL_specific_2} is interpreted as originating from the Lorentz scalar $\mathcal S\simeq \mathcal S^{(0)}= \psi^\dg \psi$.

\subsubsection{Coupling to fast oscillating field
}

With high frequency external fields described by the vector potential $\bm A(\bm r,t)$, a fast oscillating part can be neglected and only the time-independent component needs to be considered in the leading-order contribution.
Then, in the Hamiltonian, only the bilinear term with respect to the vector potential survives.
This is 
identical to the zeroth-order contribution in Floquet effective Hamiltonian theory \cite{Leskes10}.

To be more specific, we write the vector potential as
\begin{align}
    \bm A(\bm r, t) &= 
    \bm a (\bm r)\epn^{\imu \omega t} + \bm a^* (\bm r)\epn^{-\imu \omega t}
    .
\end{align}
We also impose the Coulomb gauge $\bm \nabla \cdot \bm A=0$.
We assume that the scalar potential is composed only of static components.
The electric and magnetic fields are then given by
\begin{align}
\bm E(\bm r,t) &= \bm E_0(\bm r) - \frac{\imu \omega}{c} \qty[ \bm a (\bm r) \epn^{\imu \omega t} - \bm a^* (\bm r) \epn^{\imu \omega t}  ]
,\\
    \bm B(\bm r, t) &= 
    \bm b (\bm r)\epn^{\imu \omega t} + \bm b^* (\bm r)\epn^{-\imu \omega t}
,
\end{align}
where $\bm E_0 = - \bm \nabla \Phi$
and $\bm b = \bm \nabla \times \bm a$.

Now we assume that the frequency $\omega$ is much larger than the characteristic energies in materials.
Then, we obtain the following effective Hamiltonian:
\begin{align}
    \mathscr H_{\rm eff} =& 
 \int \diff \bm r \psi^\dg \qty( 
 e\Phi+
 \frac{ \bm p^2 }{2m} 
 - \frac{\bm p^4}{8m^3c^2}
 - \frac{\hbar^2 e}{8m^2c^2} {\rm div\,} \bm E_0 
 ) \psi
\nonumber \\
&
-\frac{\hbar e}{8m^2c^2}\int \diff \bm r 
     \bm E_0 \cdot \psi^\dg  \dvec{\bm p}\times \bm \sg  \psi
\nonumber \\
 &+  \frac{\hbar^2 e^2}{8m^3c^4} \int \diff \bm r \psi^\dg \bigg( 
    \bm E_0^2 + \frac{2 \omega^2 |\bm a|^2}{c^2} - 2|\bm b|^2
    \bigg)\psi
\nonumber \\
&+\int \diff \bm r \psi^\dg \bigg(  \frac{e^2}{mc^2}|\bm a|^2
- \frac{e^4}{4m^3c^6} \qty( 2|\bm a|^4 + |\bm a^2|^2 )
\bigg) \psi
\nonumber \\
&- \frac{e^2}{m^3c^4} \int \diff \bm r  {\rm Re\,}(a_i^*a_j) p_i \psi^\dg p_j \psi
\nonumber \\
&+ \frac{\imu \hbar e^2 \omega}{2 m^2c^4} \int \diff \bm r 
 ( \bm a^* \times \bm a) \cdot \psi^\dg \bm \sg\psi
\nonumber \\
&- \frac{\hbar e^2}{2m^3c^4} \int \diff \bm r {\rm Re\,}(a_i^* b_j) \psi^\dg \dvec{p}_i \sg^j \psi 
 , \label{eq:ham_high_freq}
\end{align}
which is time-independent.
The first and second lines show the standard Pauli Hamiltonian under zero external field.
The third and fourth lines show a correction to the scalar potential.
The fifth line appears from $\bm \Pi^4$ term and gives anisotropy in the kinetic energy caused by the external field.

The sixth term shows that the high-frequency EM field generates a Zeeman-like effect.
The seventh line is also an interesting effect on electronic system: it can induce the electron chirality $\sim \bm p\cdot \bm \sg$.
Thus, the fast oscillating light can control the spin and chirality of matter.

Also in the Dirac four-component field formalism, 
the higher-order term with respect to $1/\omega$ includes the coupling between chiralities of EM field and matter.
$O(\omega^{-1})$ gives axial current coupling \cite{Ebihara16}, and $O(\omega^{-2})$ gives chirality density coupling.
The details for this aspect is summarized in Appendix C.

The quantities $\bm a^*\times \bm a$ and $a_i^*b_j$ of external field are important for the control of spin and chirality of matter.
Below, let us consider the more detailed situation based on circularly polarized light.

\begin{figure}[t]
\begin{center}
\includegraphics[width=70mm]{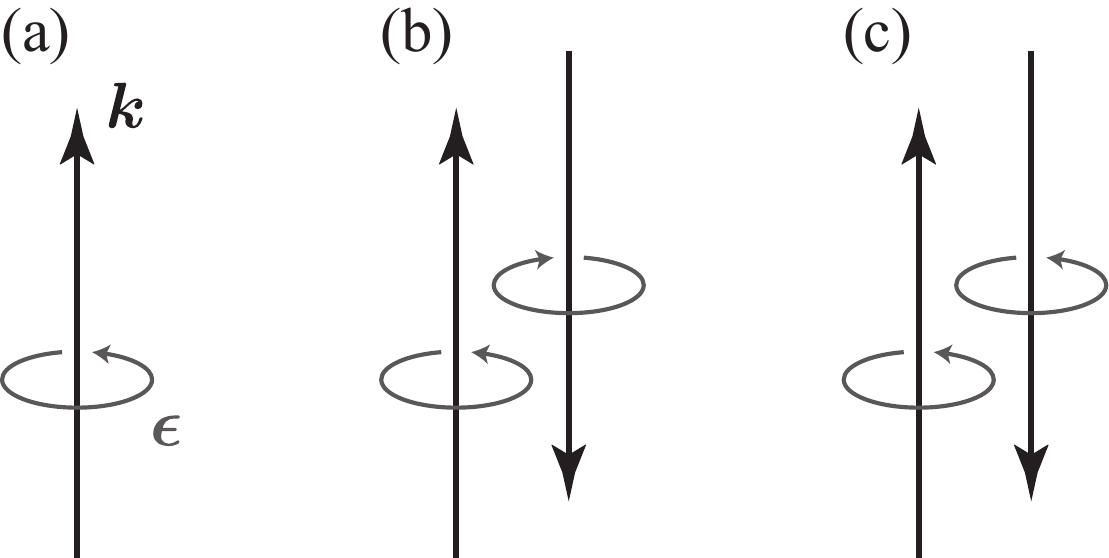}
\caption{
Schematic illustration of circularly polarized lights.
The thick arrow shows the propagating direction and the rounding arrow indicates the direction of circular polarization.
The counter-propagating lights are superposed in (b) and (c).
}
\label{fig:cpl}
\end{center}
\end{figure}

(a) {\it Simple circularly polarized light.}
\ \ \ 
The monochromatic circularly polarized light is defined by
\begin{align}
    \bm a(\bm r) &= A_0 \bm \epsilon (\hat{\bm k}) \epn^{- \imu \bm k\cdot \bm r}
    ,
\end{align}
where $\bm \epsilon$ is a polarization vector and $\bm \epsilon \times \bm \epsilon^* = -2\imu \hat {\bm k}$.
The schematic illustration is shown in Fig.~\ref{fig:cpl}(a).
For concreteness, we choose $\hat {\bm k} = \hat {\bm z}$ and $\bm \epsilon(\hat{\bm k}) = \hat{\bm x} + \imu \hat {\bm y}$.
Then we obtain 
\begin{align}
    \bm a^* \times \bm a &= 2\imu A_0^2 \hat{\bm z}
    ,\\
    a_i^* b_j &= -k a_i^* a_j = - kA_0^2 \Big[ (\delta_{ij} - \delta_{iz} \delta_{jz}) + \epsilon_{ijz} \Big]
    .
\end{align}
Namely, the circularly polarized light is a suitable {\it uniform} field for the chirality ($\bm A\cdot \bm B = {\rm const.}$).
The induction of magnetic moment by circularly polarized light can be observed as an inverse Faraday effect \cite{Kimel05}.

(b) {\it Combination of counter-propagating lights.}
\ \ \ 
Since the above simple circularly polarized light induce the magnetization and chirality simultaneously, let us also consider the following superposition of two waves with same optical helicity and amplitude [see Fig.~\ref{fig:cpl}(b)]:
\begin{align}
    \bm a(\bm r) &= A_{0} \qty( \bm \epsilon(\hat{\bm z}) \epn^{-\imu kz + \imu \phi_1} + \bm \epsilon(-\hat{\bm z}) \epn^{\imu kz +\imu \phi_2}  )
    ,
\end{align}
where $\phi_{1,2}$ are the spatially uniform phases.
The polarization vector is given by $\bm \epsilon(-\hat{\bm z}) = -\hat{\bm x} + \imu \hat{\bm y}$ \cite{Gross_book}.
Then we obtain
\begin{align}
    \bm a^* \times \bm a &= \bm 0
    ,\\
    \overline{a_i^* b_j} &= -k\overline{a_i^* a_j}= - 4kA_0^2 (\delta_{ij} - \delta_{iz} \delta_{jz})
    ,
\end{align}
where the overline indicates the spatial average which eliminates the fast oscillating part with the wavenumber $k = \omega / c$.
This combined field in principle affects the electron chirality without induction of magnetic moment (or axial current).

(c) {\it Combination of different chiralities.}
\ \ \ 
In a similar manner to the case (b), the effect on the electron chirality may be eliminated, resulting in an induction of axial current only [see Fig.~\ref{fig:cpl}(c)].
Namely, we consider 
\begin{align}
    \bm a(\bm r) &= A_{0} \qty( \bm \epsilon(\hat{\bm z}) \epn^{-\imu kz + \imu \phi_1} + \bm \epsilon^*(-\hat{\bm z}) \epn^{\imu kz +\imu \phi_2}  )
    .
\end{align}
This form indicates the superposition of the different optical helicities to cancel out the chirality effects:
\begin{align}
    \overline{\bm a^* \times \bm a} &= 4\imu A_0^2 \hat{\bm z}
    ,\\
    \overline{a_i^* b_j} &= 0
    .
\end{align}
This setup creates a uniform field conjugate to axial current, or magnetization in NRL.
Note that this effect is similar to the uniform magnetic field, but the present configuration affects only spin and does not create an orbital magnetism.

\subsection{Source term of Maxwell equation in $\bm D$-$\bm H$ representation}

Let us consider the relativistic corrections for the physical quantities associated with $\bm D$-$\bm H$ representation of the Maxwell equation. 
The charge and current densities defined by Eqs.~\eqref{eq:convective_charge} and \eqref{eq:convective_current} are given by
\begin{align}
    \rho_{\rm c} &= e\psi^\dg \psi + \frac{1}{2}\bm \nabla \cdot \bm P^{(2)} + O(m^{-3})
    ,\\
    \bm j_{\rm c} &= \frac{e}{2m} \psi^\dg \dvec{\bm \Pi} \psi 
    + O(m^{-3})  .
    \label{eq:jc_NRL_expression}
\end{align}
Although the factor $1/m$ is present in the original expression of $\rho_{\rm c}$, the time derivative $\partial_t$ produces an $O(m)$ contribution, to result in $O(1)$ for NRL.
The NRL of $\bm j_{\rm c}$ does not involve spin apparently.
In contrast, the rotation of magnetization is present for $\bm j$ given in Eq.~\eqref{eq:gordon_current}.

The relativistic corrections for the magnetic charge is given by
\begin{align}
    \rho_{\rm m} \simeq \rho_{\rm m}^{(1)} = - \bm \nabla \cdot \bm M_S 
    = e\mathcal P^{(1)}
    ,\label{eq:mag_charge_NRL}
\end{align}
where $\mathcal P^{(1)}$ is defined in Eq.~\eqref{eq:P_NRL}.
Namely, the Lorentz pseudoscalar can be interpreted as magnetic charge in $\bm D$-$\bm H$ representation.
Also, the relativistic correction for the magnetic current is given by 
\begin{align}
    \bm j_{{\rm m}}&\simeq \bm j_{{\rm m}}^{(2)} = \frac{\hbar e}{4m^2 c} \partial_i (\psi^\dg \dvec{\bm \Pi} \sg^i \psi)
    - \frac e {mc} \bm M_S \times \bm B
\end{align}
for $i=x,y,z$.
Using Eq.~\eqref{eq:spin_curr_decomposition}, it can be further rewritten as
\begin{align}
    j_{{\rm m}i}^{(2)} &= 
    \frac{\hbar e}{6m} (\bm \nabla H)_i
    + c (\bm \nabla \times \bm P_S)_i
    +
    \frac{\hbar e}{4m^2 c} \partial_j j_{Sij}^{\rm sym}
    \nonumber \\
    \ \ \ 
    &- \frac e {mc} (\bm M_S \times \bm B)_i
    .
\end{align}
This is analogous to the electric current in Eqs.~\eqref{eq:j1-2} and \eqref{eq:j2}:
for example, the terms $\bm \nabla \times \bm M_S$ and $\bm M_S \times \bm E$ in the electric current is replaced by $\bm \nabla \times \bm P_S$ and $\bm M_S\times \bm B$, respectively, in the magnetic current.
This correspondence demonstrates that $\bm j_{\rm m}$ is indeed interpreted as a magnetic version of the electric current $\bm j$.
While the rotation of the polarization, $\bm \nabla \times \bm P$, has been considered as a magnetic current \cite{Dubovik86,Hayami18}, our definition of $\bm j_{\rm m}$ stems from a pseudoscalar flow in Eq.~\eqref{eq:def_of_jm_Dirac} or Eq.~\eqref{eq:jm_orig_def}, which includes terms other than $\bm \nabla \times \bm P$.

\subsection{Quantum anomaly}

In the quantum field theory, we need to include the anomalies term in Eqs.~\eqref{eq:eom_chirality_full_relat} and \eqref{eq:eom_scalar_full_relat_anomaly}.
For the derivation, we firstly write down the equations for $\partial_t(\Psi^\dg \gm^5 \Psi) $ and $\Psi^\dg (\ \ldvec{\imu D_t}\ )\Psi$ in NRL without anomalies, and then add the anomaly terms in fully relativistic theory in Sec.~II\,F.

Combining Eq.~\eqref{eq:eom_chirality_full_relat} with Eq.~\eqref{eq:eom_NRL}, we obtain the following chirality equation of continuity:
\begin{align}
    \frac{\partial \la \tau^{Z(1)} \ra }{\partial t} + \bm \nabla \cdot \la \bm j^Z \ra \simeq \frac{2}{\hbar} \bm E\cdot \qty( \la \bm M_S \ra + \frac{e^2}{4\pi^2\hbar c} \bm B )
    , \label{eq:eom_NRL-2}
\end{align}
which is correct up to $O(m^{-2})$.
Looking at this equation, one may wonder if the $\bm E\cdot \bm B$ term creates the material chirality even in a vacuum.
This is true if we consider the particle and anti-particle creation, whose contribution is included in the original equation in Eq.~\eqref{eq:eom_chirality_full_relat}.
However, in the context of condensed matter physics, there is no antiparticle.
Then, the above relation should be regarded as the one where the matter field is present.

More specifically, in the absence of a matter field, we need to consider the following particle-antiparticle pair creation effect:
\begin{align}
    &\frac{\partial ( \epsilon_{\sg\sg'} \psi_\sg^\dg \psi_{a\sg'}^\dg+{\rm conj.} )}{\partial t} 
    + O(m^{-1})
    \nonumber \\
    &=
    \frac{2 \imu m c^2}{\hbar}  ( \epsilon_{\sg\sg'} \psi_\sg^\dg \psi_{a\sg'}^\dg - {\rm conj.} ) + 
    O(m^{-1}) + \frac{e^2}{2\pi^2\hbar^2 c} \bm E\cdot \bm B
    ,
\end{align}
where $\sg,\sg'=\ua,\da$ are spin indices.
$\psi$ and $\psi_a$ are, respectively, particle and antiparticle annihilation operator as defined in Sec.~IV [see Eq.~\eqref{eq:particle_antiparticle_operators}].
The presence of $\epsilon = \imu \sg^y$ implies the particle-antiparticle pair form a spin-singlet state, and is analogous to the standard $s$-wave superconductivity in condensed matter \cite{Schrieffer_book}.

We also discuss the relation for the Weyl anomaly in NRL given in Eq.~\eqref{eq:eom_scalar_full_relat_anomaly}.
Noting the relation in Eq.~\eqref{eq:def_of_phi_chi}, one finds that the $O(m)$ and $O(1)$ contributions exactly cancel in the Schr\"{o}dinger-field ($\psi$) representation.
The leading-order contribution is then given by 
\begin{align}
&\hbar \la \psi^\dg (\ \ldvec{\imu D_t}\ ) \psi \ra
     - \frac{1}{2m} \la \psi^\dg \ldvec{\ \bm \Pi^2} \  \psi \ra
     \nonumber \\
     &\simeq 
     - 2\bm B\cdot \la \bm M_S \ra
     - \frac{e^2}{6\pi^2\hbar c} (\bm E^2 - \bm B^2)   
     ,
\end{align}
This expression is valid up to $O(m^{-1})$.
Here again, the relation is valid only when the electrons are present since the antiparticle is removed from the theory.

\section{Bloch-Bohm theorem}

\subsection{Dirac Hamiltonian and particle/antiparticle}

The Bloch-Bohm theorem prohibits the existence of electric current in the ground state \cite{Bohm49,Ohashi96,Yamamoto15,Watanabe19}.
As shown in the previous sections, however, we have two kinds of the current densities.
One is the usual current $\bm j$ defined in terms of the Dirac field. 
The other one is $\tilde {\bm j}$, which is defined as the quantity coupled linearly to the vector potential in the non-relativistic Hamiltonian.
Since $\bm j$ and $\displaystyle \tilde {\bm j}$ are different, a question arises: can the Bloch-Bohm theorem be applied to both of the current densities?

To address this question, let us consider the ground state of condensed matter in terms of Dirac field.
To this end, we need to explicitly define the antiparticles to eliminate them from the Fock space.
In the following of this subsection, we employ the second-quantized formalism for electron field.
The Lagrangian 
is written down as
\begin{align}
\mathscr L &= \imu \int \diff \bm r \Psi^\dg \partial_t \Psi - \mathscr H
, \\
    \mathscr H &= 
    \int \diff \bm r \Psi^\dg H(A ) \Psi
    =
    \int \diff \bm r \Psi^\dg \qty(\bm \al\cdot \bm \Pi + \beta m + e\Phi)\Psi
    .
\end{align}
We take the natural unit $\hbar=c=1$.
Now we consider the unitary transformation:
\begin{align}
\Psi &= U \tilde \Psi  .
\end{align}
This is obtained by the standard 
FW transformation \cite{Foldy50,Tani49, Silenko16}, or by repeating the derivation in Sec.~III\,A in the presence of particle
and antiparticle.
Note that the definition of $\tilde \Psi$ depends on EM potentials $A^\mu(\bm r,t) = (\Phi, \bm A)$.
The transformation matrix operator is explicitly given by
\begin{align}
    U &= 1 - \frac{\beta \mathcal X}{2m} - \frac{\mathcal X^2}{8m^2} - \frac{\imu \mathcal Y}{4m^2} + \frac{3\beta \mathcal X^3}{16m^3} + \frac{\imu \beta \mathcal X \mathcal Y}{8m^3}
    + \frac{\beta \dot {\mathcal Y}}{8m^3}
    \nonumber \\
    &\ \ \ +O(m^{-4})
    ,
\end{align}
where $U^\dg = U^{-1}$ is satisfied.
We have defined 
$\mathcal X=\bm \al \cdot\bm \Pi = \begin{pmatrix} & X \\ X & \end{pmatrix}$ and $\mathcal Y = e \bm \al \cdot \bm E = \begin{pmatrix} & Y \\ Y & \end{pmatrix}$ where the $2\times 2$ matrices $X=\bm \sg \cdot \bm \Pi$ and $Y = e \bm \sg \cdot \bm E$ have been introduced in Sec.~III\,A.
Note that the transformation is not unique:
it has been pointed out that the Eriksen condition ($\beta U \beta = U^\dg$) \cite{Eriksen58,Silenko16} remove such arbitrariness, but in the present case, the condition is not satisfied due to the term proportional to $\mathcal X \mathcal Y$.
The Eriksen condition may be imposed by replacing $\mathcal X \mathcal Y$ with $[\mathcal X, \mathcal Y]/2$, but 
we stick to the above representation as it is consistent with the standard 
FW transformation \cite{Frohlich93} and the Brestetskii-Landau method in Sec. III\,A.

Now the Lagrangian is rewritten as
\begin{align}
\mathscr L &= \int \diff \bm r : \tilde \Psi^\dg (\imu \partial_t - H_{\rm eff})\tilde \Psi :
\end{align}
The colon (:) symbol represents a normal ordering.
The effective Hamiltonian is given by
\begin{align}
 H_{\rm eff} &= U^{-1}HU - \imu U^{-1}(\partial_t U)
 \\
 &\hspace{-2mm} 
 = \beta m + e\Phi + \frac{\beta \mathcal X^2}{2m} + \frac{\imu [\mathcal Y,\mathcal X]}{8m^2} - \frac{\beta \mathcal X^4}{8m^3} + \frac{\beta \mathcal Y^2}{8m^3}
 + O(m^{-4})
 .
\end{align}
Thus, the relativistic correction for the Hamiltonian is obtained in the presence of antiparticles.

Let us make a comment on the physical quantity, which is also transformed by the unitary transformation.
For example, the charge density, which is a fundamental conserved quantity, is given by
\begin{align}
    \rho &= e : \Psi^\dg \Psi : = e : ( U^* \tilde \Psi^\dg ) U\tilde \Psi: 
    \ \ \ ( \neq e : \tilde \Psi^\dg \tilde \Psi : ).
\end{align}
The inequality $\Psi^\dg \Psi \neq \tilde \Psi^\dg \tilde\Psi$ is known as a picture change error 
discussed in quantum chemistry \cite{Kello98,Mastalerz08,Fukuda16}.

We write the particle/antiparticle basis explicitly:
\begin{align}
    \tilde \Psi &= 
    \begin{pmatrix}
        \psi \\
        \epsilon \psi_a^\dg
    \end{pmatrix}
    \epn^{-\imu m t}
    \label{eq:particle_antiparticle_operators}
\end{align}
where $\epsilon = \imu \sg^y$.
The two-component spinor $\psi$ is the annihilation operator of the electron and $\psi_a$ of the positron (antiparticle).
The action is then written as 
\begin{align}
\mathscr S 
&= \int \diff^4 x (\psi^\dg \imu \partial_t \psi + \psi^\dg_a \imu \partial_t \psi_a - \psi^\dg H_p \psi - \psi_a^\dg H_a \psi_a)
,
\end{align}
where
\begin{align}
    H_p &\simeq  \hspace{9mm} 
    e\Phi + \frac{X^2}{2m} + \frac{\imu [Y,X]}{8m^2} - \frac{X^4}{8m^3} + \frac{Y^2}{8m^3}
    , \\
    H_a & \simeq
    2m 
    - e\Phi + \frac{\tilde X^2}{2m} - \frac{\imu [Y,\tilde X]}{8m^2} - \frac{\tilde X^4}{8m^3} + \frac{Y^2}{8m^3}
    ,
\end{align}
each of which corresponds to the particle and antiparticle parts.
Here, we have defined
\begin{align}
    \tilde X &= \bm \sg \cdot ( \bm p + e\bm A) .
\end{align}
As expected, the antiparticle sector is reproduced by $e \to -e$ in particle sector, which changes $X\to \tilde X$ and $Y\to - Y$.
The particle part of the Hamiltonian is equivalent to Eq.~\eqref{eq:effective_hamiltonian_up_to_order_3}.
If we want to include chemical potential for grandcanonical ensemble, as in condensed matter physics, we write the scalar potential as $e\Phi \to e\Phi - \mu$, where $\mu$ is a Lagrange multiplier to keep the total charge constant.

In a standard setup in condensed matter physics, the particle number is identical to the total charge of the electrons, which is indeed satisfied by considering $\mu$.
Namely, noting the relation $\mu \sim \ep_{\rm F}$ with the Fermi energy $\ep_{\rm F} \sim 1$ eV measured from the electron's single-particle energy bottom,
we find that the effective potential for particle is $ - \ep_{\rm F}$ (particle attractive) and for antiparticle $2mc^2 + \ep_{\rm F} \sim 10^6$ eV (antiparticle repulsive).
Hence antiparticle is energetically eliminated in condensed matter physics.

In conclusion, from the above argument, we have established the connection between the electron/positron operators
($\psi,\psi_a$) 
and the Dirac four-component field ($\Psi$) representation.

\subsection{Application of Bloch-Bohm theorem}

Now we demonstrate the Bloch-Bohm theorem for $\bm j$ and $\displaystyle\tilde {\bm j}$.
Following Ref.~\cite{Yamamoto15}, we consider the many-body ground state, which is denoted as
$|\Omega\ra$.
The Hamiltonian in terms of Dirac field is written as
\begin{align}
    \mathscr H &= \int \diff \bm r \Psi^\dg H \Psi
    .
\end{align}
Then the ground state energy is given by
\begin{align}
    E_0 &= \la \Omega | \mathscr H |\Omega\ra
    .
\end{align}
We now consider the trial state
\begin{align}
|\Omega' \ra &= \epn^{\imu \delta \bm p \cdot \bm x} |\Omega\ra ,
\\
    {\bm x} &= \int \diff \bm r \Psi^\dg \bm r \Psi ,
\end{align}
where we have introduced the positional operator $\bm x$.
Assuming that $|\delta \bm p|$ is sufficiently small,  we obtain
\begin{align}
    \la \Omega'|\mathscr H | \Omega' \ra
    &= E_0 + \delta \bm p \cdot \la \Omega | \int \diff \bm r \Psi^\dg \bm \al \Psi |\Omega \ra
    + O(\delta \bm p^2) ,
\end{align}
where we have used the anticommutation relation for the fermionic field operator $\Psi$.
The energy of the trial state $|\Omega'\ra$ cannot be lower than the ground state energy $E_0$ according to the Rayleigh-Ritz theorem, which results in the conclusion that the total current must be zero: $\int \diff \bm r \la \Omega| \bm j | \Omega\ra= 0$.

However, one ambiguity is present in the above argument. Namely, the negative infinity energy of the Dirac sea is present without explicit normal ordering, with which the ground state energy cannot be defined. 
The finite ground state energy must be evaluated by considering normal ordering for antiparticle operators. 
With this in mind,
let us consider another form of the Hamiltonian after unitary transformation:
\begin{align}
    \tilde {\mathscr H} &=  \int \diff \bm r \, :\tilde \Psi^\dg H_{\rm eff}\tilde \Psi :
    = \int \diff \bm r (\psi^\dg H_p  \psi + \psi_a^\dg H_{a}  \psi_a) ,
\end{align}
where the normal ordering is explicitly performed.

Now we consider the ground state composed of particles only, denoted as $|\Omega_p\ra$.
Its energy is given by $E_{0p}$, which is finite and is well-defined.
We also introduce the trial state
\begin{align}
|\Omega_p' \ra &= \epn^{\imu \delta \bm p \cdot \bm x_p} |\Omega_p\ra
,\\
    {\bm x}_p &= \int \diff \bm r \psi^\dg \bm r \psi
\end{align}
where the position operator $\bm x_p$ is defined with the Schr\"odinger field.
The energy is calculated as
\begin{align}
    \la \Omega_p' | \mathscr H |\Omega_p'\ra
    &= E_{0p} + \imu \delta \bm p \cdot \int \diff \bm r \la \Omega | \psi^\dg[ H_p, \bm r ] \psi |\Omega\ra  .
\end{align}
We explicitly evaluate the commutation relation in the $m\to \infty$ limit:
\begin{align}
    [H_p,\bm r] &= - \frac{\imu }{m} \bm \Pi + \frac{\imu e}{4m^2} \bm \sg \times \bm E 
    +O(m^{-3})
    .
\end{align}
This calculation shows that the vanishing current is $\int\diff \bm r \la \Omega_p| \tilde {\bm j} |\Omega_p\ra$ defined in Eq.~\eqref{eq:def_of_tilde_current}, not the true electric current $\int \diff \bm r \la \Omega_p| {\bm j} |\Omega_p\ra$.

Actually, 
the position operator can be substituted by another operator, resulting in a different physical quantity becoming zero.
Let us try the following generalized (or relativistic) position operator \cite{Zou20, Ado24}
\begin{align}
    \bm X &= \int \diff \bm r \psi^\dg \qty( \bm r + \frac{1}{4m^2}\bm \Pi\times \bm \sg + O(m^{-3}) )  \psi .
\end{align}
It is notable that $\bm \Pi \times \bm \sg = \bm p\times \bm \sg - e \bm A\times \bm \sg$ serves as a position operator in relativistic quantum mechanics \cite{Foldy50,Zou20}.
 Repeating the previous argument, we arrive at the conclusion of vanishing $\int \diff \bm r \la \Omega_p| {\bm j} |\Omega_p\ra$ up to $O(m^{-2})$ in the ground state composed only of particles.
 The disappearance of $\int \diff \bm r \la \Omega_p| {\bm j} |\Omega_p\ra$ should persist to higher orders in $1/m$ expansion if we properly choose the relativistic position operator.
Thus, while the current density ($\tilde {\bm j}$) defined by the derivative of effective Hamiltonian with respect to the vector potential $\bm A$ has a different form with the current density ($\bm j$) defined by the original Dirac four-component field, both the quantities become generally zero in the ground state.
The no-go-theorem is thus confirmed for $\displaystyle \tilde {\bm j}$ and $\bm j$.

\begin{table*}[t]
 \caption{
List of the leading-order relativistic corrections for the Dirac bilinears and their conjugate EM fields.
Note that the expressions for relativistic correction are not fully listed in this table, and the coefficients are omitted (see the main text for the complete expressions).
The listed quantities are all gauge invariant and Hermitian.
}
  \begin{tabular}{ccccc}
\hline
Physical Quantity & Symbol & \ \ (SI/TR)\ \ \  & 
NRL and 
Relativistic Corrections &  Conjugate Field 
\\
\hline
Charge
& $\rho = e \bar \Psi \gm^0 \Psi $ & ($+/+$) & $\psi^\dg \psi $ , $\bm \nabla \cdot \bm P_S $ & $ \Phi$
\\[1mm]
Current
& $\bm j = ec \bar \Psi \bm \gm \Psi $ & ($-/-$) & $ \psi^\dg \dvec{\bm \Pi} \psi$ , $ \bm \nabla\times \bm M_S $
& $ \bm A$
\\[1mm]
Magnetization
& $\bm M = \frac{\hbar e}{2mc} \bar \Psi \bm \Sigma \Psi$ & ($+/-$) & 
$\bm M_S =\frac{\hbar e}{2mc} \psi^\dg \bm \sg \psi $
& $ \bm B$
\\[1mm]
Electric Polarization & $\bm P =\frac{\hbar e}{2mc} \bar \Psi ( - \imu \bm \al ) \Psi$ & ($-/+$) & $\bm \nabla (\psi^\dg \psi)$
, $\bm P_S = \frac{\hbar e}{8m^2c^2} \psi^\dg \dvec{\bm \Pi}\times \bm \sg \psi $ & $ \bm E $
\\[1mm]
Chirality & $\mathcal A^0 = -\tau^Z = \bar \Psi \gm^0 \gm^5 \Psi$ & ($-/+$) & 
 $H=\frac{1}{2mc} \psi^\dg \dvec{\bm \Pi}\cdot  \bm \sg \psi $ & $ \bm A\cdot \bm B$
\\[1mm]
Axial Current & $ \bm{\mathcal A}= \bar \Psi \bm \gm \gm^5 \Psi $ & ($+/-$) &  
$ \bm M_S$, 
$ \bm j^Z = \frac{1}{4m^2 c} [\psi^\dg \dvec{\bm \Pi} \big( \dvec{\bm \Pi}\cdot \bm \sg \big) \psi
+\frac \hbar 2 \bm \nabla \times \psi^\dg \dvec{\bm \Pi} \psi]
$ & $\bm E \times \bm A $
\\[1mm]
Lorentz Pseudoscalar & $\mathcal P = \tau^Y = \bar \Psi \imu \gm^5 \Psi$ & ($-/-$) & 
 $ \bm \nabla \cdot \bm M_S$ & $ \bm E\cdot \bm B$
\\[1mm]
Lorentz Scalar & $\mathcal S =\tau^X = \bar \Psi \Psi$ & ($+/+$) & $\psi^\dg \psi $ , $\bm \nabla \cdot \bm P_S $ & $ m$ (mass), $\bm E^2 - \bm B^2$
\\[1mm]
\hline
Chirality Polarization & $\bm {\mathcal P}_C = \frac c 2 \bar \Psi \dvec{\bm \Pi}\times \bm \gm \Psi 
$ & ($+/+$) &  $\bm \nabla H$, $\psi^\dg \ldvec{(\imu \bm \Pi^2)}\bm \sg \psi$ & N/A
\\[1mm]
Magnetic Charge & $\rho_{\rm m} = - \bm \nabla \cdot \bm M$ & ($-/-$) &  $\bm \nabla \cdot \bm M_S$ & N/A
\\[1mm]
Magnetic Current & $\bm j_{\rm m} =-\frac{e}{2m} \bar \Psi \dvec{\bm \Pi} \imu \gm^5 \Psi 
$ & ($+/+$) &  $\bm \nabla H,\ \bm \nabla \times \bm P_S$ & N/A
\\[1mm]
\hline
  \end{tabular}
\label{tab:classification}
\end{table*}

\section{Summary and Discussion}

We have clarified the microscopic physical quantities in condensed matter physics derived from the Dirac field in relativistic quantum mechanics.
The results are summarized in Tab.~\ref{tab:classification}.
We also show that the EM field can control the charge and chirality of electrons.
These physical quantities are defined in {\it any} materials and can be evaluated using first principles calculation, which quantifies the characteristics of materials.
The accumulated knowledge based on systematic evaluation for each material will lead to a search for the useful functionalities.

For example, the polarity and chirality of the system are characterized by $\int \diff \bm r \la\bm P_S\ra $ and $\int \diff \bm r \la \tau^Z  \ra$.
The actual evaluation in materials is shown as a separate publication \cite{Miki24}.
Since the spin is involved in these expressions, the spin-orbit coupling (SOC) is necessary to obtain the finite value.
Hence, the large value is expected if the material includes heavy elements.
On the other hand, the value of e.g. $\int \diff \bm r \la \tau^Z  \ra$ can be utilized even for weak-SOC materials.
The magnitude of these quantities is expected to be dependent linearly on the SOC energy for light elements.
Then, if these quantities are normalized by the SOC energy, it can characterize the polarity and chirality of weak-SOC materials and we can compare the values for different materials.
We note that the chirality can also be quantitatively characterized by the hydrodynamic helicity $\bm j\cdot (\bm \nabla \times \bm j) $, which is defined as a quantum operator as shown in Sec.~III.

As shown in Tab.~\ref{tab:classification}, 
the Dirac bilinears are categorized by the symmetry of SI/TR and (one-component)scalar/(three-component)vector.
In this regard, the scalar with (SI/TR)$=(+/-)$ is still missing in the table.
Noting that the scalar can be constructed by taking the divergence of the vector as in $\rho_{\rm m} = - \bm \nabla \cdot \bm M$ (see the row of `Magnetic Charge'), the desired scalar is obtained as $\bm \nabla \cdot \bm j$ which indeed has the symmetry (SI/TR)$=(+/-)$.
The physical meaning of this scalar is clear from the equation of continuity, as it is identical to $- \partial_t \rho$.
We remark that the source term for the Lorentz scalar $ \mathcal S$, i.e., the right-hand side $\bar \Psi \dvec{\bm \Pi} \cdot (-\imu \bm \al) \Psi$ of Eq.~\eqref{eq:scalar_eom}, also serves to determine a ${\rm (SI/TR)=(+/-)}$ scalar.
Note also that these scalars $\partial_t\rho$ and $\partial_t \mathcal S$ are absent in stationary state.

The present paper just systematically clarifies and classifies the microscopic physical quantities based on the relativistic quantum mechanics.
Their concrete expectation values in condensed matter remain to be explored.
From a wider perspective, the relativistic corrections exist also in the interactions \cite{Breit29, Ito65, Hoshino23Breit}.
Actually, the content of this paper is limited to the case where the EM field is regarded as an external classical field, while the quantum treatment involving photons, which mediate the interaction between electrons, complicates the analysis.
Moreover, a specific experimental method for observing the physical quantities remains an open question.
A systematic study of the relativistic effect for condensed matter both theoretically and experimentally will open a new avenue for intriguing many-body physics.

\section*{Acknowledgement}
SH thanks M. Senami, S. Ozaki, M. Fukuda, M.J.S. Yang, and S. Michimura for their valuable discussions. 
He also expresses his gratitude to H. Aoki and R. Arita for insightful conversations on Floquet theory, and to Y. Tanii for critical reading of our manuscript.
This work was supported by the KAKENHI Grants No.~23K25827 (SH,MTS,HI), No.~21K03459 (SH), No.~24K00578 (SH), No.~23KJ0298 (TM), No.~24K00588 (MTS), No.~24K00581 (MTS).

\appendix

\section{Chiral and Weyl anomalies in statistical mechanics}

Following Ref.~\cite{Fujikawa_book}, we summarize the derivation of the quantum anomalies in statistical mechanics.
The equations are derived using imaginary time, a formulation that is convenient for condensed matter physicists. 
The real-time representation can then be obtained via Wick rotation.
This procedure is similar to the derivation of the equation of motion for spins \cite{Nagaosa_book, Auerbach_book}.

We begin with the imaginary-time action for the Dirac electrons ($\hbar=c=k_{\rm B}=1$):
\begin{align}
    S [\Psi] &= \int_0^{1/T} \diff \tau \int \diff \bm r 
    \, \Psi^\dg (\partial_\tau + \bm \al \cdot \bm\Pi + \beta m + e\Phi) \Psi
    .
\end{align}
Here, $\Psi$ is a four component vector of Grassmann numbers.
The upper limit of $\tau$-integral is the inverse temperature $1/T$.
We also define the gamma matrices $\gm_{\rm E}^0 = \imu\beta$ and
 $\gm_{\rm E}^i  =\gm^i$ in Euclidean space.
 The action is then rewritten as
\begin{align}
    S 
&= \int\diff^4 x 
    \, \bar \Psi  \qty( -\gm_{\rm E}^{0}(\partial_0 + \imu e A_0) - \gm^i (\partial_i + \imu e A_i) - \imu m) 
    \nonumber \\
&= \imu \int\diff^4 x 
    \, \bar \Psi  \qty( \imu \slashed D - m)     
\Psi
    ,
\end{align}
where $\slashed{D} = \gm_{\rm E}^\mu D_\mu = \gm_{\rm E}^\mu (\partial_\mu + \imu e A_\mu)$ is an invariant derivative and $A^0 = - A_0 = \imu \Phi$ is the imaginary scalar potential.
The Euclidean metric is introduced by $g_{\mu\nu} = g^{\mu\nu} = {\rm diag\,}(-1,-1,-1,-1)$.
The Dirac conjugate is given by $\bar \Psi = \Psi^\dg \gm_{\rm E}^0$.

Now we consider the general infinitesimal transformation $\Psi' = \epn^{\imu \lambda(x) X}\Psi$ where $x^\mu = (\tau,\bm r)$ ($\mu = 0,1,2,3$) where $\lambda(x) \in \mathbb C$.
We assume that $X$ is Hermitian ($X^\dg = X$), and obtain
\begin{widetext}
\begin{align}
    S [\Psi]
    &= S[\Psi'] - \imu \int \diff^4 x \ {\rm Re\,}\lambda(x) \Big\{
    -\partial_\tau ({\Psi'}^\dg X{\Psi'})
    + \frac 1 2 \imu \partial_i ({\Psi'}^\dg \{ \al^i,X\} {\Psi'}) 
    + \frac 1 2 {\Psi'}^\dg \dvec{\Pi}_i [\al^i,X] {\Psi'}
    + m {\Psi'}^\dg [ \beta, X] {\Psi'}
    \Big\}
    \nonumber \\
    &\hspace{14mm}
    + \int \diff^4 x \ {\rm Im\,}\lambda(x) 
    \Big\{
     {\Psi'}^\dg \dvec{D}_\tau X\Psi' 
     + \frac 1 2 {\Psi'}^\dg \dvec{\Pi}_i \{\al^i,X\} \Psi'
     + \frac 1 2 \imu \partial_i ( {\Psi'}^\dg  [\al^i,X] \Psi' )
     + m {\Psi'}^\dg \{\beta,X\} \Psi'
    \Big\}  ,
\end{align}
where $i=1,2,3$ (or $x,y,z$) is the index for space components.
$D_\tau = \partial_\tau + e\Phi$ is a gauge-invariant imaginary-time derivative.

Next, let us consider the change in Jacobian in the transformation of the integral measure \cite{Fujikawa_book}.
We expand the Dirac field and its conjugate as
\begin{align}
    \Psi(x) &= \sum_n a_n \varphi_n(x) ,
    \\
    \bar \Psi (x) &= \sum_n \tilde b_n \varphi^\dg_n(x) ,
\end{align}
where $a,b$ are Grassmann numbers and $\varphi_n (x) = \la x|n\ra$ is a four-component-vector basis functions.
The transformed version is given by
\begin{align}
    \Psi'(x) &= \sum_n a_n' \varphi_n(x)
    = \sum_n a_n [ \varphi_n(x) + \imu \lambda(x)X \varphi_n(x) ] 
    , \\
    \bar \Psi' (x) &= \sum_n \tilde b_n' \varphi^\dg_n(x)
    = \sum_n \tilde b_n [ \varphi^\dg_n(x) - \imu \lambda^*(x) \varphi^\dg_n(x) \gm_{\rm E}^{0\dg} X \gm_{E}^0 ] 
    .
\end{align}
The path-integral measure is then transformed as
\begin{align}
    \mathscr D \tilde b' \mathscr D a'
    &= \det \qty(\delta_{nn'} + \imu \int \diff x  \varphi_n^\dg (x) \qty[ \lambda(x) X - \lambda^*(x)\gm_{\rm E}^{0\dg} X \gm_{\rm E}^0 ] \varphi_{n'}(x) )^{-1} 
\mathscr D \tilde b
    \mathscr D a
    .
\end{align}
The partition function is given by the path-integral of the Boltzmann factor:
\begin{align}
         &Z = \int \mathscr D \bar \Psi \mathscr D \Psi \, \epn^{-S}
         .\label{eq:part_func}
\end{align}
Noting $\mathscr D \bar \Psi \mathscr D \Psi =
\mathscr D \bar b
    \mathscr D a$, we obtain the change in the action due to the Fujikawa's Jacobian as
\begin{align}
    S_\lambda &= - \imu \int \diff^4 x \sum_n \varphi_n^\dg (x) \Big[ {\rm Re\,}\lambda(x) Y_- + \imu \, {\rm Im\,}\lambda(x) Y_+  \Big] \varphi_n(x)
    ,
\end{align}
where $Y_\pm = X \pm \gm_{\rm E}^{0\dg} X \gm_{\rm E}^0$.
We have used
$\det (1+X) \simeq \det \epn^X = \exp ({\rm Tr\,} X)$.
For the evaluation of the integral,
the gauge-invariant regularization is employed:
\begin{align}
        S_\lambda &\to - \imu \int \diff^4 x \sum_{n} \varphi_n^\dg(x) \Big[ {\rm Re\,}\lambda(x) Y_- + \imu \, {\rm Im\,}\lambda(x) Y_+  \Big] f(\slashed{D}^2/M^2) \varphi_n(x)
        ,
\end{align}
where $f(t) = \epn^{-t}$ and $M$ will be taken to be infinity.

Now we use the concrete basis $\varphi_n (x) = \frac{1}{\sqrt{V/T}} \epn^{-\imu k x} u_p$ with $n = (k,p)$ and $kx = k^\mu x_\mu = - (\bm k\cdot \bm r + \omega_m \tau)$.
$u_p$ is a four-component orthonormal vector basis.
$k^i = 2\pi n_i/L$ is a wavenumber with the system volume $V=L^3$, and $\omega_m = (2m+1)\pi T$ is fermionic Matsubara frequency ($n_i,m \in \mathbb Z$).
Then the action is given by
\begin{align}
    S_\lambda &= - \imu \int \diff^4 x \  \frac{1}{V/T}\sum_{k} \epn^{\imu kx} {\rm tr\,} \bigg\{ \Big[ {\rm Re\,}\lambda(x) Y_- + \imu \, {\rm Im\,}\lambda(x) Y_+  \Big] f(\slashed{D}^2/M^2) \bigg\} \epn^{-\imu kx}
    .
\end{align}
We utilize the following relations:
\begin{align}
    \slashed D^2 &= D_\mu D^\mu  + \frac{\imu e}{4} [\gm_{\rm E}^\mu,\gm_{\rm E}^\nu] F_{\mu\nu}
    , \\
    F_{\mu\nu} &= \partial_\mu A_\nu - \partial_\nu A_\mu
    ,
\end{align}
where we have used $\{ \gm_{\rm E}^\mu, \gm_{\rm E}^\nu\} = 2g^{\mu\nu}$.
Thus we evaluate the action as
\begin{align}
    S_\lambda
    &= - \imu\int \diff^4 x \  \lambda(x) \frac{1}{V/T}\sum_{\bm kn} {\rm tr\,} \Big[ {\rm Re\,}\lambda(x) Y_- + \imu \, {\rm Im\,}\lambda(x) Y_+  \Big] f\qty( \frac{(\imu k+D)^2}{M^2}  - \frac{e}{M^2} (F_{0i}\al^i + B_i \Sigma^i) )
    .
\end{align}
We define $F_{ij} = - \epsilon_{ijk} B_k$ ($\bm B$ is the magnetic field).
The term without EM field is neglected because of an unconnected contribution \cite{Fujikawa_book}.
In the $M\to \infty$ limit, 
we evaluate the Gaussian integrals such as 
\begin{align}
    &\lim_{M\to \infty} \frac{1}{M^4 V/T} \sum_{n_1n_2n_3m} \epn^{k^2/M^2} 
    =\frac{1}{16\pi^2}
    .
\end{align}
and then obtain
\begin{align}
    S_\lambda &= - \frac{\imu e^2}{32\pi^2} \int \diff^4 x \  {\rm tr\,}  \Big[ {\rm Re\,}\lambda(x) Y_- + \imu \, {\rm Im\,}\lambda(x) Y_+  \Big]  \qty[ \frac 2 3 (F_{0i}^2+B_i^2)1 -2 F_{0i}B_i \gm^5 ]
    .
\end{align}
To derive this relation, we have implicitly assumed $X=1$ or $\gm^5$.
The Schwinger-Dyson equation is thus completed:
by choosing $X=\gm^5$ ($Y_+ = 0$ and $Y_-=2\gm^5$), we obtain
\begin{align}
&    \partial_\tau \la \Psi^\dg \gm^5\Psi \ra
    - \frac 1 2 \imu \partial_i \la \Psi^\dg \{ \al^i,\gm^5\} \Psi \ra
    - m \la  \Psi^\dg [ \beta, \gm^5] \Psi  \ra
    + \frac{e^2}{2\pi^2} F_{0i} B_i
    = 0
    ,
\end{align}
and by choosing $X=1$ ($Y_+ = 2$ and $Y_- = 0$),
\begin{align}
\la {\Psi}^\dg \dvec{D}_\tau \Psi  \ra
     +  \la {\Psi}^\dg \dvec{\Pi}_i \al^i \Psi \ra
     + 2 m \la {\Psi}^\dg \beta \Psi \ra
     + \frac{e^2}{6\pi^2} (F_{0i}^2 + B_i^2) = 0
     .
\end{align}
\end{widetext}
The expectation value is defined by $\la \cdots\ra = \int \mathscr D \bar \Psi \mathscr D \Psi (\cdots )\epn^{-S} / Z$ where the partition function is defined by Eq.~\eqref{eq:part_func}.
Now we move to the Minkowski space by Wick rotation: $\tau \to \imu t$ and 
$F_{0i} \to - \imu E_i$ ($\bm E$ is the electric field).
Then, we obtain
\begin{align}
&\partial_t \la \bar \Psi \gm^0 \gm^5 \Psi \ra
+ \bm \nabla \cdot \la \bar \Psi \bm \gm \gm^5 \Psi\ra
    = 2m \la \bar \Psi (\imu \gm^5) \Psi \ra
    - \frac{e^2}{2\pi^2} \bm E\cdot \bm B
    ,\label{eq:anom_chiral}
    \\
& \la {\bar \Psi} \gm^0 \ldvec{(\imu D_t)} \Psi \ra
     -  \la {\bar \Psi}\bm \gm \cdot \dvec{\bm \Pi} \Psi \ra
     =
     2 m \la {\bar \Psi} \Psi \ra
     - \frac{e^2}{6\pi^2} (\bm E^2 - \bm B^2)
     .\label{eq:anom_weyl}
\end{align}
The right-hand sides include quantum anomaly terms: the chiral anomaly in Eq.~\eqref{eq:anom_chiral} and the Weyl anomaly in Eq.~\eqref{eq:anom_weyl} \cite{Fujikawa_book}.
Note the relation $\bar \Psi\gm^0 \gm^5 \Psi = \psi_L^\dg \psi_L - \psi_R^\dg \psi_R$ in our definition.

Specifically for the Weyl anomaly, the equation is written in terms of the energy-momentum tensor 
\cite{Peskin_book}.
Since 
a part of Eq.~\eqref{eq:anom_weyl} is regarded as
the trace of energy-momentum tensor, the relation is also known as the trace anomaly.

\section{Electron gas model with chirality and polarity}

We consider the simple electron gas model
\begin{align}
    \mathscr H &= \int \diff \bm r \, \psi^\dg
    \qty( \frac{\bm \Pi^2}{2m}
    -\mu
    -\frac{\hbar e}{2mc} \bm B \cdot \bm \sg
    + \lambda_{ij} \Pi_i \sg^j
    ) \psi
    ,
\end{align}
which is spatially uniform in the absence of the magnetic field.
Note that the spin-orbit coupling $\lambda_{ij}$ has a dimension of velocity.
We write it as
\begin{align}
    \lambda_{ij} &= \frac{\lambda}{3}\delta_{ij} + \frac 1 2 \epsilon_{ijk} \eta_k
    ,\label{eq:def_of_lamij}
\end{align}
where the parameters $\lambda$ and $\bm \eta$ characterize chiral and polar electronic systems, respectively.
We assume that the spin-orbit coupling is small compared to the Fermi energy, and expand physical quantities with respect to $\lambda_{ij}$.

\subsection{Spin current tensor}

Let us first consider the expectation value of the spin current density in the absence of the external field ($\bm A=\bm 0$):
\begin{align}
    \la j_{Sij} \ra&=\la \psi^\dg \dvec{p}_i \sg^j \psi\ra 
    \simeq \frac{4\hbar^2 I_2}{3}  \lambda_{ij}
    , \label{eq:spin_current_gas_model}
\end{align}
where
\begin{align}
    I_2 &= 
    \frac{1}{V/T} \sum_{m\bm k} \frac{\bm k^2}{(\imu\omega_m - \xi_{\bm k})^2}
    \\
    &= -\frac{2m}{\hbar^2} D_0 \ep_{\rm F}
    < 0.
\end{align}
The last expression of $I_2$ is obtained at zero temperature ($T = 0$).
Here we have defined $\xi_{\bm k} = \frac{\hbar^2 \bm k^2}{2m} - \mu$.
We note the relations $\mu=\ep_{\rm F}$ at $T=0$, $k_{\rm F}^3 = 3 \pi^2 n$ ($n$ is the number density), and $D_0 \ep_{\rm F} = \frac{3n}{4}$ ($D_0$ is the density of states at Fermi level) for a degenerated ideal Fermi gas with spin $1/2$.

At zero temperature, we obtain
\begin{align}
    \la H\ra &\simeq - n \frac{\lambda}{c}
    ,\label{eq:appendix_hel_value}
    \\
    \la \bm P_{S} \ra &\simeq 
    - \frac 1 4 \cdot \frac{\hbar}{mc} \cdot en \cdot \frac{\bm \eta}{c}  ,
\end{align}
where the parameters in the right-hand sides are defined in Eq.~\eqref{eq:def_of_lamij}.
Note that $H$ has a dimension of number density, and $\bm P_S$ has the dimension same as a familiar electric polarization $\bm P(\bm r) = \bm r\rho(\bm r)$.
Assuming that the typical spin-orbit coupling is $\lambda_{ij} \sim 0.1 v_{\rm F}$ with the Fermi velocity $v_{\rm F} =\hbar k_{\rm F}/m \sim 0.01 c$, we obtain
$\frac{H}{n} \sim 10^{-3}$.
Namely, the difference between right-handed and left-handed electrons is roughly 0.1\% for a typical spin-orbit coupled metal.

\subsection{Hydrodynamical helicity}

Next, we consider the quantum contribution to the hydrodynamical helicity
defined by Eq.~\eqref{eq:hydro_helicity} in our simple electron gas model.
The Fourier representation of the helicity is given by
\begin{align}
    \mathcal H_{\rm hydro} &=
    \frac 1 V \sum_{\bm q} : \bm j(-\bm q) \cdot \Big[ \imu \bm q\times \bm j(\bm q) \Big] :
\end{align}
where $\bm j(\bm q) = \int \diff \bm r \bm j(\bm r)\epn^{-\imu \bm q\cdot \bm r}$.
The colon symbol (:) represents the normal ordering, which ensures that the two-body quantity is well-defined since the single-particle state does not contribute.
The current density is introduced based on the Hamiltonian as $\bm j(\bm r) = - c \frac{\delta \mathscr H}{\delta \bm A (\bm r)}$.
The electric current in $\bm q$-space is then given by
\begin{widetext}
\begin{align}
    j_i(\bm q)
    &= e \sum_{\bm k} c_{\bm k}^\dg
    \qty[ \frac{\hbar}{2m} (2k_i +q_i)\hat 1  +  \qty( \lambda_{ij} - \frac{\imu \hbar }{2m} \epsilon_{ijk} q_k )\hat \sg^j  ]
    c_{\bm k+\bm q}
    ,
\end{align}
where $c_{\bm k} = \frac{1}{\sqrt V} \int \diff \bm r \psi(\bm r) \epn^{-\imu \bm k\cdot \bm r}$ is the two-component spinor (annihilation operator) with wavevector $\bm k$.
The expectation value of the hydrodynamical helicity is then given by
\begin{align}
    &\hspace{-7mm}  \la \mathcal H_{\rm hydro}\ra \simeq - \frac{\imu e^2}{V} \sum_{\bm k \bm k'} \epsilon_{ijk} (k'_i-k_i) 
    \ {\rm Tr\,} 
    \qty[ \frac{\hbar}{2m} (k'_j + k_j) \hat 1 + \qty(\lambda_{jl_1} - \imu \frac{\hbar}{2m} \epsilon_{jl_1m_1}(k'_{m_1}-k_{m_1}))\hat \sg^{l_1} ]
\qty[ 
    f(\xi_{\bm k'}) \hat 1 + f'(\xi_{\bm k'}) \hbar k'_{p_2} \lambda_{p_2q_2}\hat \sg^{q_2} ]
    \nonumber \\
    &\hspace{30mm} \times 
    \qty[ \frac{\hbar}{2m} (k'_k + k_k) \hat 1 + \qty(\lambda_{kl_2} + \imu \frac{\hbar}{2m} \epsilon_{kl_2m_2}(k'_{m_2}-k_{m_2}))\hat \sg^{l_2} ]
    \qty[ 
    f(\xi_{\bm k}) \hat 1 + f'(\xi_{\bm k}) \hbar k_{p_1} \lambda_{p_1q_1}\hat \sg^{q_1} ]
    ,
\end{align}
where we have expanded the expression with respect to the parameter $\lambda$.
We have introduced the Fermi distribution function by $\displaystyle f(x) = T\sum_m \frac{\epn^{\imu \omega_m 0^+}}{\imu\omega_m - x} = 1/ (\epn^{\beta x}+1)$.
It is easily confirmed that $\lambda=0$ gives zero helicity.
Keeping first-order terms with respect to $\lambda$, 
and using the formulae such as
\begin{align}
   \frac{1}{V} \sum_{\bm k} k_i k_j k_l k_m g(\xi_{\bm k})
    = \frac{\delta_{ij} \delta_{lm} + \delta_{il} \delta_{jm} + \delta_{im} \delta_{jl}}{15}  \cdot \frac{1}{V} \sum_{\bm k} \bm k^4 g(\xi_{\bm k}) 
\end{align}
for an arbitrary function $g$,
we obtain
\begin{align}
    \la \mathcal H_{\rm hydro} \ra &\simeq 
     - \frac{4\hbar }{V} \lambda \qty(\frac{\hbar e}{2m})^2 \sum_{\bm k \bm k'} f'(\xi_{\bm k'})  f(\xi_{\bm k})
     \qty(  {\bm k'}^4  + \frac{25}{9}\bm k^2 {\bm k'}^2  )
    -   
    \frac{8e}{V} \lambda\qty(\frac{\hbar e}{2m})\sum_{\bm k \bm k'} 
    f(\xi_{\bm k'})
    f(\xi_{\bm k})
    \frac{2\bm k^2}{3} .
\end{align}
\end{widetext}
The wavevector integration at $T=0$ gives
\begin{align}
    \la \mathcal H_{\rm hydro}\ra &\simeq  \frac{49}{10} \cdot V (en v_{\rm F})^2 k_{\rm F} \cdot \frac{\lambda}{v_{\rm F}}
    .
\end{align}
This expression except for the numerical factor can be understood by the dimensional analysis for the ideal electron gas.
The hydrodynamical helicity is connected to the electron chirality density through the relation \eqref{eq:appendix_hel_value}.

\section{High frequency expansion of effective Hamiltonian}
Here, we extend the discussion in Ref.~\cite{Ebihara16} and derive the coupling between chiralities of electrons and EM fields.
We consider the time-periodic Hamiltonian $\mathscr H(t)$, and define the time-independent operator $\mathscr H^{[n]}$ by
\begin{align}
    &\mathscr H(t) = \sum_n \mathscr H^{[n]} \epn^{\imu n\omega t}
    ,
\end{align}
where the period is given by $T=\omega/2\pi$.
The superscript number with square bracket specifies each Fourier component.
We consider the concrete form of the Dirac Hamiltonian in the second-quantization form
\begin{align}
    \mathscr H(t) &= \int \diff \bm r \Psi^\dg \qty[ c \bm \al \cdot \qty(\bm p - \frac e c \bm A(t)) + \beta mc^2 + e\Phi ]\Psi ,
    \\
    \bm A(\bm r,t)  &= \bm a(\bm r) \epn^{\imu \omega t} + \bm a^*(\bm r) \epn^{- \imu \omega t} ,
\end{align}
which includes only single mode with (high) frequency $\omega$.
Then
\begin{align}
\mathscr H^{[0]} &= \int \diff \bm r \Psi^\dg \qty[ -\imu \hbar c \bm \al \cdot \bm \nabla + \beta mc^2 + e\Phi ]\Psi
    ,\\
    \mathscr H^{[n\neq 0]} &= - e \int \diff \bm r \Psi^\dg \bm \al \Psi \cdot \Big[ \delta_{n1} \bm a(\bm r) + \delta_{n,-1} \bm a^*(\bm r) \Big]  ,
\end{align}
where only $n=-1,0,1$ components are included in the Hamiltonian.

Now we construct the high-frequency expansion of the Floquet effective Hamiltonian \cite{Leskes10}. 
This is valid for $\hbar \omega \gg 2mc^2$.
Hence, the effective Hamiltonian based on $\mathscr H^{(0,1,2,\cdots)}$ (the number in parenthesis indicates the order of $1/m$) discussed in Sec.~III, which is derived for non-relativistic limit with an implicit assumption of $mc^2 \gg \hbar \omega$, is different from the following expressions.
Noting the magnitude of rest mass energy $mc^2\simeq 0.511$MeV, the experimentally relevant light in this context may be a gamma ray, which is not frequently encountered in the field of condensed matter physics.

The first-order effective Hamiltonian is trivial: $\mathscr H_1 = \mathscr H^{[0]}$ where the subscript number indicates the order of $\mathscr H^{[n]}$.
The second-order effective Hamiltonian at large $\omega$ is \cite{VanVleck29,Leskes10} 
\begin{align}
    \mathscr H_{2} &= \frac 1 2 \sum_{n\neq 0} \frac{[\mathscr H^{[n]} , \mathscr H^{[-n]} ]}{n\hbar \omega}
    .
\end{align}
This is the essentially same as those obtained in a Floquet theory of condensed matter \cite{Oka09,Kitagawa11,Lindner11,Wang14,Mikami16}.
By evaluating the commutation relation, we obtain \cite{Ebihara16}
\begin{align}
    \mathscr H_2 
    &= - \frac{2\imu e^2}{\hbar\omega} \int \diff \bm r [\bm a(\bm r) \times \bm a(\bm r)^*] \cdot \bar \Psi \bm \gm \gm^5 \Psi .
\end{align}
The conjugate field to the axial current is thus identified.

We also consider the higher-order terms.
The third-order effective Hamiltonian (second-order in $1/\omega$) is given for a single-mode case 
by \cite{Leskes10,Mikami16} 
\begin{align}
    \mathscr H_3 &= - \frac 1 3 \sum_{n,n'\neq 0, n\neq n'} \frac{[[\mathscr H^{[n-n']},\mathscr H^{[n']}],\mathscr H^{[-n]}]}{nn'(\hbar\omega)^2}
    \nonumber \\
&    - \frac 1 2 \sum_{n\neq 0} \frac{[[\mathscr H^{[0]},\mathscr H^{[n]}],\mathscr H^{[-n]}]}{n^2(\hbar\omega)^2}
.
\end{align}
For the Hamiltonian discussed in this section, we have
\begin{align}
    \mathscr H_3 &= 
    - \frac{1}{2(\hbar\omega)^2} \sum_{n=\pm 1} \qty[ [\mathscr H^{[0]},\mathscr H^{[n]}],\mathscr H^{[-n]}]
     = \mathscr H_3^\dg 
     .
\end{align}
After straightforward calculation, we obtain
\begin{align}
    & \frac{-2(\hbar\omega)^2}{e^2} \mathscr H_3 = 
4\hbar c 
\int \diff \bm r 
\big[ {\rm Re\,}\bm a^* \cdot (\bm \nabla \times \bm a)  \big]
\Psi^\dg \gm^5 \Psi
\nonumber \\
&\ \ \ + 8 
\int \diff \bm r 
\qty( |\bm a|^2\delta_{ij} - {\rm Re\,}a_i^* a_j ) \Pi_{ij}
+ 8mc^2 \int \diff \bm r |\bm a|^2 \Psi^\dg \gm^0 \Psi
.
\end{align}
We have defined the 
kinetic energy
tensor by
\begin{align}
    \Pi_{ij} &= \frac 1 2\Psi^\dg \dvec{p_i} (c\al^j) \Psi
    .
\end{align}
Noting the relations
\begin{align}
\la A^i(\bm r,t)A^j(\bm r,t) \ra &= 2 {\rm Re\,} a_i^{*} a_j
    ,\\
    \bm A(\bm r,t)\times \bm E(\bm r,t) 
    &= \frac{2\imu \omega}{c} \bm a\times \bm a^*
,\\
    \la \bm A(\bm r,t)\cdot \bm B(\bm r,t) \ra
    &=  2 {\rm Re\,} \bm a^* \cdot (\bm \nabla \times \bm a)
    ,
\end{align}
where the square bracket indicates the time average $\la \cdots\ra = \int_0^T  (\cdots ) \frac{\diff t}{T}$ ($T=2\pi/\omega$),
we can 
write down the chirality part as
\begin{align}
    \mathscr H_{\rm ext}' &= - \frac{e^2 c}{\hbar\omega^2} \int \diff \bm r \big(
    \la \bm A\cdot \bm B \ra \bar \Psi \gm^0 \gm^5 \Psi
    + \la \bm A\times \bm E \ra \cdot  \bar \Psi \bm \gm \gm^5 \Psi
 \big) .
\end{align}
We may rewrite it as
\begin{align}
    \mathscr H'_{\rm ext} &= - \frac{e^2 c}{\hbar \omega^2}\int \diff \bm r  \, \la \tilde A_\mu \ra \, \bar\Psi \gm^\mu \gm^5 \Psi ,
\end{align}
where $\tilde A^\mu$ is defined in Eq.~\eqref{eq:def_of_tilde_A}.
Clearly, this relation is analogous to the standard coupling between electron and EM potentials:
\begin{align}
    \mathscr H_{\rm ext}
    &= - e\int \diff \bm r \, A_\mu \, \bar\Psi \gm^\mu \Psi ,
\end{align}
where the coupling constant is $e$.
Hence, $\la \tilde A_\mu \ra = (\la \bm A\cdot \bm B\ra, \la \bm A\times \bm E\ra)$ 
are, respectively, chiral scalar potential and chiral vector potential, where the coupling constant is given by $e^2 c / \hbar \omega^2$ for the both cases.

\end{document}